\documentclass[a4paper,12pt]{article}
\pdfoutput=1
\usepackage{jheppub}
\usepackage{graphicx}
\usepackage{amssymb}
\usepackage{amsmath}
\usepackage{amsfonts}
\usepackage{hyperref}
\usepackage{xcolor}

	\def\bea{\begin{eqnarray}}
	\def\eea{\end{eqnarray}}
	\def\be{\begin{equation}}
	\def\ee{\end{equation}}
	\def\nn{\nonumber\\}

	\def\mm{\mathcal}
	
	\def\pa{\partial}
	\def\zb{{\bar{z}}}
	\def\wb{{\bar{w}}}

	\def\rz{\rho_z}
	\def\rzb{{\bar{\rho}_z}}
	\def\rw{\rho_w}
	\def\rwb{{\bar{\rho}_w}}
	
	\definecolor{db}{rgb}{0,0.08,0.45}
	\definecolor{brick}{rgb}{0.6,0.1,0.3}
	
	\definecolor{zz}{rgb}{1,0,0}
	\definecolor{zz2}{rgb}{0.7,0.1,0.1}
	\definecolor{yy}{rgb}{0.05,0.9,0.05}
	\definecolor{ww}{rgb}{0.6,0.1,0.3}
	
	\definecolor{rr}{cmyk}{0,0,0,1}
	\definecolor{vv}{rgb}{0.5,0,0.5}
	\definecolor{ss}{cmyk}{0,0,0,1}

	\definecolor{brick}{rgb}{0.5,0,0.5}
	
\def\Res{{\rm Res}}

	 \def\b{\beta}  \def\d{\delta}     \def\m{\mu}  \def\r{\rho}  \def\k{\kappa}   
	\def\G{\Gamma}  \def\D{\Delta}

\title{A Conformal Dispersion Relation:\\ Correlations from Absorption}

\author[\Psi]{Dean Carmi}
\author[\sigma]{Simon Caron-Huot}
\affiliation[\Psi]{Institute of Physics, \'Ecole Polytechnique F\'ed\'erale de Lausanne (EPFL), Rte de la Sorge, BSP 728, CH-1015 Lausanne, Switzerland}
\affiliation[\sigma]{Department of Physics, McGill University, 3600 Rue University, Montr\'eal, QC Canada H3A 2T8}
\emailAdd{deancarmi1@gmail.com}
\emailAdd{schuot@physics.mcgill.ca}

\abstract{
We introduce the analog of Kramers-Kronig dispersion relations for correlators of four scalar operators
in an arbitrary conformal field theory.
The correlator is expressed as an integral over its ``absorptive part",
defined as a double discontinuity, times a theory-independent kernel which we compute explicitly.
The kernel is found by resumming the data obtained by the Lorentzian inversion formula.
For scalars of equal scaling dimensions, it is a remarkably simple function (elliptic integral function) of two pairs of cross-ratios. We perform various checks of the dispersion relation (generalized free fields, holographic theories at tree-level, 3D Ising model), and get perfect matching.
Finally, we derive an integral relation that relates the ``inverted" conformal block with the ordinary conformal block.
}

\begin{document} 
\maketitle
\flushbottom

\section{Introduction}

The conformal bootstrap has enjoyed remarkable success in the last decade, employing both numerical
\cite{Rattazzi:2008pe,ElShowk:2012ht,Simmons-Duffin:2016wlq,Poland:2018epd} and analytic \cite{Komargodski:2012ek,Fitzpatrick:2012yx,Alday:2015ewa,Alday:2016njk} methods to solve general consistency conditions. Some of the primary methods of the analytic bootstrap include: light-cone expansions of the crossing equations, large $N$ expansions, AdS/CFT, and causality constraints. 

Implications of causality are often effectively captured by dispersion relations, following the work
of Kramers and Kronig in optics. These authors related (in 1926)
the dispersive (real) and absorptive (imaginary) part of the index of refraction,
exploiting analyticity of the index of refraction in the upper-half complex frequency plane.
Dispersion relations were later used to try and constrain the relativistic S-matrix \cite{Martin:1969ina,Eden1,Screaton}.
This was an important tool for physicists in the 1950's and 60's who, in the absence of a microscopic theory, attempted to solve or ``bootstrap'' the strong interactions using  consistency with the principles of causality, unitarity, and crossing;
a program which waned down at the time with the advent of QCD as a microscopic description of the strong force.

Dispersion relations are typically most useful when one knows more about the absorptive part than the real part.
For the strong force at low energies, the imaginary part is often saturated  by narrow resonances,
leading to phenomenologically interesting sum rules \cite{Shifman:1992xu}.
It may also happen that the imaginary part (or the absolute value of the amplitude) is the only quantity measured experimentally.
Theoretically, the imaginary part enjoys useful properties such as positivity (for example in the forward limit),
related to probabilities being nonnegative; applications include
the first proof of irreversibility of renormalization group flow in four spacetime dimensions \cite{Komargodski:2011vj}.
In perturbative scattering amplitudes, absorptive parts can be efficiently computed in terms of lower-order amplitudes through the Cutkosky rules, a foundational insight
that is now built into successful methods such as generalized unitarity
\cite{Bern:1994zx,Britto:2004nc,Britto:2005fq,Elvang:2013cua}.
Given that crossing symmetry and general principles appear to be particularly powerful in conformal field theories,
it is natural to expect a CFT dispersion relation to be a useful tool in constraining CFT correlators.

In this paper we derive a dispersion relation for CFT 4-point correlators $\mm{G}(z,\zb)$:
\be
\label{eq:main1z}
\boxed{
\mathcal{G}^t(z,\zb) = \int_0^1 dw d\wb K(z, \zb, w , \wb) {\rm dDisc}[\mm{G}(w,\wb)]}
\ee 
where we separate the $t$ and $u$ channel contributions and a possible finite sum of non-normalizable blocks (see section \ref{sec:subtleties}):
\be
\mm{G}(z,\zb) = \mm{G}^t(z,\zb) +\mm{G}^u(z,\zb) + \mbox{(non-norm.)}\,.   \label{G from Gtu}
\ee 
The input ${\rm dDisc}[\mm{G}(z,\zb)] $ represents the double-discontinuity of the correlator around $\zb =1$,
defined below, and is interpreted physically as its absorptive part.
We notice that the correlator is a function of two cross-ratios $(z,\zb)$:
the kernel $K(z, \zb, w , \wb)$ is thus a function of two \emph{pairs} of cross-ratios, one pair being integrated over (with $w,\wb$ real in the integration region).
This is to be contrasted with more familiar Kramers-Kronig type dispersion relations, in which a single variable is integrated over.
We will argue that such a complication is unavoidable if we insist that the input be the ``absorptive part" ${\rm dDisc}[\mm{G}]$,
as the analytic properties of the correlators $\mm{G}(z,\zb)$ entangle its two arguments.

The existence of a formula such as (\ref{eq:main1z}), reconstructing
correlators from (double) discontinuities, is suggested by the Lorentzian inversion formula of
\cite{Caron-Huot:2017vep,Simmons-Duffin:2017nub,Kravchuk:2018htv}.
That formula reconstructs operator product expansion data from knowledge of the discontinuities
${\rm dDisc}[\mm{G}(z,\zb)]$ of the CFT 4-point correlator, and has been used notably
to streamline light-cone and large-$N$ expansions.
Examples suggest that a crude approximation to the ${\rm dDisc}$ 
({\it ie.} including the simplest few exchanged operators) can lead to accurate results to the OPE data itself.
These examples range from the low-twist spectrum in 3D Ising and related models
\cite{Alday:2017zzv,Alday:2019clp,Albayrak:2019gnz,Cardona:2018qrt,Li:2019dix}, mean field theory \cite{Liu:2018jhs},
the calculation of Witten diagrams in strongly coupled (holographic) gauge theories \cite{Alday:2017vkk,Caron-Huot:2018kta},
as well as defect CFTs and certain finite temperature effects \cite{Lemos:2017vnx,Iliesiu:2018fao}.

We find it is extremely encouraging that good first approximations to the dDisc are easy to come by.
This begs the question of systematic improvement. One limitation of the Lorentzian inversion formula
is that it is difficult to iterate it.  For example, its output cannot simply be fed back into it,
in a way that would lead to successively better approximations
(while the formula produces a generating function for the spectrum, computing the dDisc requires resolving
the dimensions of individual operators, a step which requires a numerically difficult analytic continuation).
The dispersion relation (\ref{eq:main1z}) offers a step forward,
since it enables crossing equations to be formulated directly on the positive dDisc.
As we will see, it will also circumvent technical limitations regarding convergence at low spins.

In this paper we derive the dispersion relation (\ref{eq:main1z}), and in particular the kernel $K$ entering it,
by resumming the OPE data extracted via the Lorentzian inversion formula.
The result can be split into a two-dimensional \textit{bulk integral} $K_B$ and a one-dimensional \textit{contact integral} $K_C$\footnote{The $\r$-variables, defined in eq.~(\ref{rho}), is: $\rz \equiv \frac{1-\sqrt{1-z}}{1+\sqrt{1-z}}$, and similarly for $\zb, w, \wb$.}:
\be \label{KBC}
K(z, \zb, w , \wb) = K_B \theta(\rz\rzb\rwb-\rw) + K_C \frac{d\rw}{dw}\delta( \rw - \rz\rzb\rwb)
\ee
where $\theta(x)$ is the unit step function and $\delta(x)$ is the Dirac $\delta$-function.
In the case of operators of equal external scaling dimensions, our main result is the explicit form:
\begin{align}  \label{eq:pol67}
\begin{aligned}
K_B&= -\frac{1}{64\pi} \left(\frac{z\zb}{w\wb}\right)^{3/2}
\frac{(\wb -w)(\frac{1}{w}+\frac{1}{\wb}+\frac{1}{z}+\frac{1}{\zb}-2)}{((1-z)(1-\zb)(1-w)(1-\wb))^{\frac{3}{4}}}
\ x^{\frac{3}{2}}  {}_2 F_1 (\tfrac{1}{2},\tfrac{3}{2},2,1-x),
\\
K_C &=\frac{4}{\pi}  \frac{1}{\wb^2}%\frac{\sqrt{1-\rz^2\r_\zb^2 \r_\wb^2}}{\sqrt{(1-\rz^2)(1-\r_\zb^2)(1-\r_\wb^2)}}
\left( \frac{1-\rz^2\rzb^2 \rwb^2}{(1-\rz^2)(1-\rzb^2)(1-\rwb^2)}\right)^{1/2}
\frac{1-\rz\rzb\rwb^2}{(1-\rz\rwb)(1-\rzb\rwb)}.
\end{aligned}
\end{align}
The first involves a rather special combination of cross ratios:
\be \label{x}
x \equiv \frac{\rz\rzb\rw\rwb(1-\rz^2)(1-\rzb^2)(1-\rw^2)(1-\rwb^2)}{(\rzb\rwb- \rw \rz)(\rz\rwb- \rw \rzb)(\rz\rzb -\rw\rwb)(1-\rw \rz \rwb\rzb)} .
\ee 
The bulk integral contributes only for $\rw < \rz \rwb \rzb$ (due to the step function),
and is proportional to a hypergeometric function, which can equivalently be written as a combination of elliptic integral functions, see eq.~(\ref{eq:elliptic2}).
The contact integral, proportional to a $\delta$-function, is effectively integrated over a single variable $\wb\in [0,1]$.
An alternative but equivalent form, which unites the bulk and contact terms, is given in eq.~(\ref{G combined}).

We find it remarkable that a function of four complex variable can be written in closed form as in 
eq.~(\ref{eq:pol67}).  As we will see in section \ref{sec:directproof}, each factor plays a role, and $K$ above is arguably the simplest
possible kernel able to fulfil the difficult task assigned to it.

The outline of the paper is as follows. In section~\ref{sec:amp2} we review the amplitude dispersion relation and the Froissart-Gribov inversion formula, and how one can derive the former from the latter.
This exercise will prepare us for the more difficult case of the CFT dispersion relation. In section~\ref{sec:disp} we show the full details of derivation of the CFT dispersion relation in $d=2$ for scalars with equal external scaling dimensions. We obtain an analytic result for the kernel, in terms of elliptic integral functions.
The same kernel is valid in any dimension, and we show that in section~\ref{sec:4d} that indeed repeating the calculation in $d=4$ yields the same kernel.
In section~\ref{sec:unequal} we derive the dispersion relation for unequal external scaling dimension. The kernel satisfies a differential equation, giving Taylor expansions for it. For a specific simple case, $a=0$ and $b=\frac{1}{2}$, we also find an analytic form for the kernel. In section~\ref{sec:directproof} we establish the validity of the
dispersion relation by a direct contour deformation argument. This allows to overcome some of the original assumptions,
and in particular we obtain a subtracted dispersion relation that is valid in any unitary CFT.
In section~\ref{sec:checks} we explore possible applications of the dispersion relation:
to strong coupling $\mm N =4$ SYM,
to obtain novel identities relating inverted and conformal blocks,
and to the 3D Ising model and new bootstrap functionals.
We conclude by discussing future directions in section~\ref{sec:discussion}.
\\

\noindent {\it Note added}:  While this paper was being completed,
the work \cite{Bissi:2019kkx} appeared on arxiv who introduced a single-variable dispersion relation
that reconstructs correlators from a single-discontinuity. This appears to be quite distinct from the formulas considered here:
the input in this case (to our knowledge) is neither sign-definite nor admits a physical interpretation as an absorptive part.

%%%%%%%%%%%%%%%%%%%%%%%%%%%%%%%%%%%%%%%%%%%%%%%%%%%%%%%%

\section{Preliminaries}
\label{sec:review}

\subsection{Review of amplitude dispersion relation}

\label{sec:amp2}

Dispersion relations enable to construct a function from a knowledge of it's discontinuities.
The most common type of a dispersion relation is the single variable dispersion relation, where
one variable is being integrated over. For definiteness, we will discuss this here in the context of the relativistic 4-particle scattering amplitude, although the reader may wish to keep in mind that the construction is more general.
We will review two derivations, the first involving a contour deformation argument which is perhaps the most familiar.

\begin{figure}[!h]
	\centering
	\includegraphics[width=135mm]{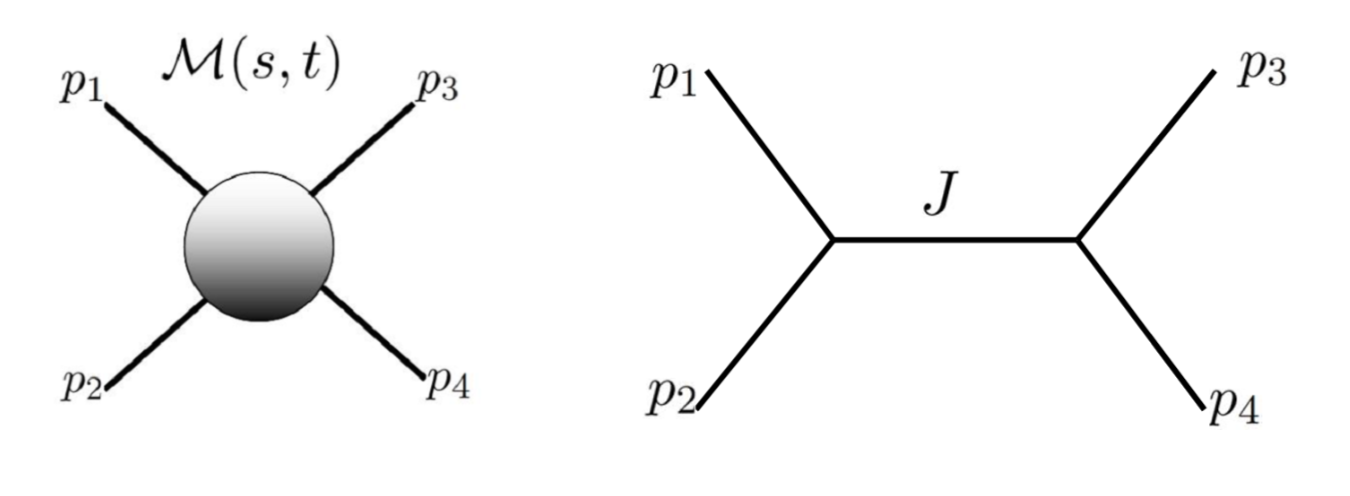}
	\caption{Left: The 4-particle scattering amplitude $\mm{M}(s,t)$, with external momenta $p_i$. Right: $t$-channel tree level exchange diagram of particle with spin $J$ \label{f4}.}
\end{figure}

\begin{figure}[!h]
	\centering
\be\begin{array}{cc}
\hspace{0mm}\def\svgwidth{68mm}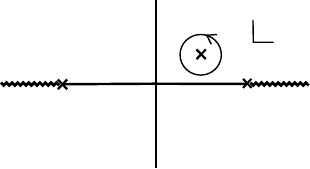
& \hspace{6mm}
\raisebox{17mm}{\resizebox{7mm}{!}{$\Rightarrow$}}
\hspace{6mm}\def\svgwidth{68mm}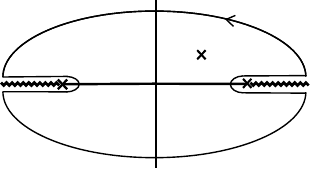
\end{array}\nonumber\ee
	\caption{Left: The amplitude can be written as a contour integral by using Cauchy's theorem. Right: Upon deforming the contour, there will be contributions from the branch cuts and from the arcs at infinity. \label{f1}}
\end{figure}

Consider the 4-particle scattering amplitude $\mm{M}(s,t)$ for scalars with mass $m$  (fig.~\ref{f4} left).
$\mm{M}(s,t)$ is a function of the two Mandelstam variables $s\equiv -(p_1+p_2)^2$ and $t\equiv -(p_1-p_3)^2$,
with the energy conservation constraint $s+t+u= 4m^2$.
For $s$ constant and in a suitable range, the complex $t$-plane has the structure depicted in fig.~\ref{f1},
with two branch cuts along the
real axis for $t>t_0$ and $t<4m^2-s-t_0$. These are called the $s$- and $t$-channel cuts (the second condition
corresponding to $u>u_0$). The single variable dispersion relation to be considered is:
\bea \label{amp_dispersion_relation}
\mm{M}(s,t)= \frac{1}{2\pi}\int_{t_0}^\infty \frac{dt'}{t'-t} {\rm Disc}_{t'}[\mm{M}(s,t')]+ (t \leftrightarrow u)
\eea 
The integral runs over the branch cuts of $\mm{M}(s,t')$, and $i\ {\rm Disc}_{t'}[\mm{M}(s,t')] \equiv  \mm{M}(s,t'+i0)- \mm{M}(s,t'-i0)$ is the discontinuity across the cuts in the $t'$-plane. Note that the variable $s$ just goes along for the ride\footnote{One could alternatively write a dispersion relation in the $s$-plane, with fixed $t$.}.

A common way to derive this is to start with a contour integral in the complex $t'$
plane surrounding the point $t$ (see Fig~\ref{f1}); by Cauchy's residue theorem: 
\be
\label{eq:pol0}
\mm{M}(s,t) = \frac{1}{2\pi i}\oint \frac{dt'}{t'-t}  \mm{M}(s,t') .
\ee
Then one deforms the contour of integration as in Fig~\ref{f1}.
If $\mm{M}(s,t')$ decays fast enough at $|t'| \to \infty$ that the arcs at infinity can be neglected,
only the branch cuts contribute, reproducing eq.~(\ref{amp_dispersion_relation}) as desired.
%\footnote{If the amplitude grows polynomially, one can perform a subtractions procedure for the dispersion relation.}.

What if $\mm{M}(s,t')$ does not decay fast enough?
If it is polynomially bounded, one can still obtain a subtracted dispersion relation.
The idea is to improve the behavior on large arcs by subtracting the amplitude at some reference $t=t_*$:
\be \label{once subtracted}
 \mm{M}(s,t) - \mm{M}(s,t_*) =\frac{1}{2\pi}
 \int_{t_0}^\infty dt' \left[ \frac{1}{t'-t} - \frac{1}{t'-t_*}\right] 
%\frac{dt' (t-t_*)}{(t'-t)(t'-t_*)}
 {\rm Disc}_{t'}[\mm{M}(s,t')]  + \mbox{u-channel},
\ee
which has improved convergence since the bracket $\sim 1/|t'|^2$.  One can generalize by applying more subtractions as needed.
A perhaps more illuminating way to write this is to use
elementary algebra to rewrite the bracket as  $\frac{t-t_*}{(t'-t)(t'-t_*)}$, and divide both sides by $(t-t_*)$;
the once-subtracted dispersion relation (\ref{once subtracted}) becomes:
\be \label{once subtracted 1}
\frac{\mm{M}(s,t)}{t-t_*} = 
\frac{1}{2\pi}\int_{t_0}^\infty \frac{dt'}{t'-t}\  {\rm Disc}_{t'}\left[ \frac{\mm{M}(s,t')}{t'-t_*}\right]
+\mbox{u-channel}.
\ee
This is nothing but the original dispersion relation, now applied to the rescaled function\footnote{In this form, we have assumed
that $t_*$ is inside the integration range, so that the discontinuity contains a term $-2\pi\delta(t'-t_*) \mm{M}(s,t_*)$.}
$\mm{M}(s,t)/(t-t_*)$.
This viewpoint will be useful below.

In the amplitude context, the dispersion integral generally runs over unphysical regions of the $(s,t')$-plane,
where ${\rm Disc}_{t'}\mm{M}$ is neither positive-definite nor physically measurable.
An exception is for the range $0\leq s<4m^2$ in a theory with mass gap $m$:
there the discontinuity \emph{is} positive-definite, and is a smooth extrapolation
(to imaginary angles) of physically measurable $t$- and $u$-channel scattering amplitudes.
This is an important result of Martin,
used in his celebrated proof of the Froissard bound on the high-energy growth of total cross-sections \cite{Martin:1962rt}.
The CFT dispersion relation discussed in this paper will share the nice features of this special region.

%Mandelstam proposed a double dispersion relation for the scattering amplitude:   % not sure...
%\bea \mm{M}(s,t) = \int ds dt \frac{\r_{s't'}}{(s'-s)(t'-t)} +\int ds du \frac{\r_{s'u'}}{(s'-s)(u'-u)}+\int du' dt' \frac{\r_{ut}}{(u'-u)(t'-t)} \eea
%Where $2\pi i \r_{s't'} \equiv Disc_{s'}Disc_{t'} \mm{M}(s,t) $

For more on applications of scattering amplitude dispersion relations, the reader may consult \cite{Martin:1969ina,Eden1,Screaton}.
For a more recent application, see the following works on the S-matrix bootstrap \cite{Paulos:2016but,Paulos:2017fhb}.

%%%%%%%%%%%%%%%%%%%%%%%%%%%%%%%%%%%%%%%%%%%%%%%%%%%%%%%%%%%%

\subsubsection{Dispersion relation from the Froissart-Gribov formula}
\label{ssec:amp1}

We turn to a perhaps less familiar derivation of the dispersion relation, starting from the Froissart-Gribov formula
expressing partial wave coefficients from the discontinuity of the amplitude.

Consider the partial wave decomposition of the amplitude in the $s$-channel (Fig.~\ref{f4} right),
for definiteness working in $d=4$ dimensions:
\be
\label{eq:p00}
\mm{M}(s,t(z)) = \frac{1}{2}\sum_{J=0}^\infty (2J+1) a_J (s) P_J(z),  \qquad z\equiv \cos \theta, \qquad  t(z)= \frac{4m^2-s}{2}(1-z).
\ee
Physically, $\theta$ is the scattering angle and the coefficients $a_J(s)$ encode the decomposition of the amplitude
into spherical harmonics at a given energy-squared $s$.

Using the orthogonality of the Legendre polynomials, $\int_{-1}^1  dz P_{J'}(z)P_J(z)= \frac{2\d_{JJ'}}{2J+1}$,
one may readily obtain a ``Euclidean inversion formula'' expressing the coefficients as an integral over the amplitude.
A less obvious formula, first derived by Froissart and Gribov \cite{Froissart,Gribov}, expresses the same data
in terms of the \emph{discontinuity} of the amplitude:
\be
\label{eq:p0}
a_J^t(s) = \frac{1}{\pi} \int_{1}^\infty dz'  Q_J(z')\ {\rm Disc}_{t} \mm{M}(s,t(z')), \qquad a_J(s) = a_J^t(s) + (-1)^J a^u_J(s),
\ee 
where $Q_J(z_s')$ is the Legendre function of the second kind.  $a_J^t(s)$ and $a_J^u(s)$ are the contributions from the $t$-channel and $u$-channel cuts respectively.
The Frossart-Gribov formula plays a foundational role in Regge theory, as it establishes analyticity in spin of the partial waves
(as well as providing quantitative large-spin estimates).

A proof of eq.~(\ref{eq:p0}) starts from the orthogonality relation, 
rewriting the integral over $z\in [-1,1]$ as a contour integral using that $P_J \propto {\rm Disc}\ Q_J$.
One then deforms the contour exactly as in fig.~\ref{f1} above
(see \cite{Caron-Huot:2017vep} for recent discussion with two derivations).
The Froissart-Gribov formula and dispersion relation are thus closely related, and it should come as no surprise that
one can derive either one from the other.

To go the other way, the trick is simply to plug the coefficient obtained from 
eq.~(\ref{eq:p0}) into the partial wave sum in eq.~(\ref{eq:p00}),
and interchange the summation and integration:
\bea
\label{eq:p000}
\mm{M}(s,t(z)) = \sum_{J=0}^\infty (2J+1)  P_J(z)  \frac{1}{2\pi}\int_1^\infty  dz'  Q_J(z') {\rm Disc}_{t} \mm{M}(s,t(z')) + (t \leftrightarrow u)
\nn
 = \frac{1}{2\pi}\int_{1}^\infty  dz'  {\rm Disc}_{t} \mm{M}(s,t(z'))   \sum_{J=0}^\infty (2J+1)  P_J(z)  Q_J(z') + (t \leftrightarrow u).
\eea
The latter sum then turns into the following identity (for $|z|<|z'|$):
\footnote{This can be proved by combining the following two equations:
\be  \frac{1}{2} \sum_{J=0}^{\infty} (2J+1)P_J(z)P_J(z') = \d(z-z'), \quad\mbox{and}\quad
Q_J(z)= \frac{1}{2}\int_{-1}^{1} \frac{dz' P_J(z')}{z'-z}.
\ee
The latter shows that $P_J(z)$ equals the discontinuity across the cut of $Q_J(z)$.}
\be
\label{eq:dvd14}
 \sum_{J=0}^\infty (2J+1)  P_J(z)  Q_J(z') = \frac{1}{z'-z}
\ee
which is recognized as the kernel of the dispersion relation (\ref{amp_dispersion_relation}).
(One needs only the change of variable $\frac{dz'}{z'-z} \mapsto \frac{dt'}{t'-t}$.)

We call the measure $\frac{dt'}{t'-t}$, which multiplies the discontinuity, the ``kernel".
Interestingly, even though the form of the special functions $P_J$ and $Q_J$ changes in a complicated way
as a function of spacetime dimension, and the left-hand-side of eq.~(\ref{eq:dvd14}) acquires a measure factor
$[(1-z'^2)/(1-z)^2]^{(d-4)/2}$, one can show that the sum produces the same right-hand-side in any dimension.
This was to be expected physically, since the dimension simply did not enter the earlier derivation anywhere.

The reader may wonder why one would want to derive a dispersion relation starting from the Froissart-Gribov formula (\ref{eq:p0}),
as opposed to simply writing down the more elementary Cauchy kernel $\frac{dt'}{t'-t}$.
The reason is that the substitution $P_J\mapsto {\rm Disc}\ Q_J$ underlying the former
has a group-theoretical explanation ({\it ie.} both functions satisfy the same Casimir differential equation),
whereas writing down the Cauchy kernel requires an educated guess.
For conformal correlators, the group-theoretic approach
was successfully carried out in ref.~\cite{Caron-Huot:2017vep},
whereas the guessing approach turns out to be much more challenging.

\subsection{Review of CFT kinematics}

In this paper we will focus on a correlator of four scalar primary operators in a CFT.
This can be written as a function of cross ratios $z$ and $\zb$ multiplied by an overall factor which is determined by the conformal symmetry:
\be \label{stripped correlator}
\langle  \mm{O}_1(x_1)  \mm{O}_2(x_2) \mm{O}_3(x_3) \mm{O}_4(x_4)   \rangle =\frac{\Big( \frac{x^2_{14}}{x^2_{24}} \Big)^{a} \Big( \frac{x^2_{14}}{x^2_{13}} \Big)^{b}}{(x^2_{12})^{\frac{1}{2}(\D_1 +\D_2)}(x^2_{34})^{\frac{1}{2}(\D_3 +\D_4)}} \mm{G}(z,\zb)\,,
\ee 
where we defined the differences of the external scaling dimensions:
\be
a= \tfrac{1}{2}(\D_2 -\D_1), \qquad b= \tfrac{1}{2}(\D_3 -\D_4)\,
\ee 
and the cross ratios $z,\zb$ are defined through:
\be
u= z\zb= \frac{x^2_{12}x^2_{34}}{x^2_{13}x^2_{24}}\ , \qquad v= (1-z)(1-\zb)= \frac{x^2_{23}x^2_{14}}{x^2_{13}x^2_{24}}\ .
\ee
We will often use the so-called radial or $\r$-coordinates of ref.~\cite{Hogervorst:2013sma},
\be \label{rho}
\rz \equiv \frac{1-\sqrt{1-z}}{1+\sqrt{1-z}} \ , \quad 
\rzb\equiv \frac{1-\sqrt{1-\zb}}{1+\sqrt{1-\zb}}\ , \quad z=\frac{4\rz}{(1+\rz )^2}\ , \quad \zb=\frac{4\rzb}{(1+\rzb)^2}
\ee
which provide a double cover of the complex $z$-plane.

We will be focusing on the $s$-channel operator product expansion (OPE):
\bea
\mm{G}(z,\zb) = \sum_{J,\D}  f_{12\mm{O}_{J,\D}} f_{43\mm{O}_{J,\D}} G_{J,\D}(z,\zb)
\eea 
where $ f_{ij\mm{O}}$ are the OPE coefficients and $G_{J,\D}(z, \zb)$ are $s$-channel conformal blocks
for exchange of a primary operator with spin $J$ and scaling dimension $\D$, and its descendants.
For our purposes the OPE may also be written as an integral over principal series representations (harmonic functions),
in which the scaling dimension is continuous (see \cite{Dobrev:1975ru,Dobrev:1977qv,Costa:2012cb}):
\be
\label{eq:p1}
\mm{G}(z,\zb) =\sum_{J=0}^{\infty}\int_{\frac{d}{2}- i \infty}^{\frac{d}{2}+ i \infty} \frac{d \D}{2\pi i} c(J,\D) F_{J,\D}(z,\zb) + \mbox{(non-norm.)}.
\ee 
The ``non-normalizable" part includes the $s$-channel identity operator as well as a possible finite sum of $F$ functions for
scalar operators with dimension less than $d/2$.
The CFT data is then encoded in the poles of $c_{J,\D}$, which occur on the real axis of the complex $\D$ plane at the position of the physical scaling dimensions, and whose residue are the squared OPE coefficients:
\be
f_{12\mm{O}_{J,\D}} f_{43\mm{O}_{J,\D}}   =  - \Res_{\D'=\D} \big[ c(J,\D') \big].
\ee 
The $F$ stand for harmonic functions, which combine a block and its shadow
\be
F_{J,\D}(z,\zb) = \frac{1}{2} \Big(  G_{J,\D}(z,\zb)  + \# G_{J,d-\D}(z,\zb)  \Big),
\ee 
with a specific coefficient that will not be important below. (It ensures
that $F$ is single-valued in Euclidean space where $\zb=z^*$, 
a necessary condition for the $F$'s to form a complete orthogonal basis.)

Using the orthogonality for $F_{J,\D}(z,\zb )$, one may readily write an Euclidean inversion formula expressing
the OPE data  $c_{J,\D}$ as an integral over correlators,
in analogy to that for Legendre polynomials discussed below eq.~(\ref{eq:p00}).
Instead we will use the Lorentzian inversion formula, which reconstructs the same data
from an ``absorptive part" \cite{Caron-Huot:2017vep,Simmons-Duffin:2017nub,Kravchuk:2018htv}
\be
\label{eq:p2}
c^t(J,\D) = \frac{\k_{J+\D}}{4} \int_0^1 dw d\wb\ \m(w,\wb)
\ G_{\D+1-d,J+d-1}(w,\wb)\ {\rm dDisc}[\mm{G}(w,\wb)] 
\ee 
where the integration region is the square $ 0 \leq w , \wb \leq 1$, the normalization and measure are
\be \label{kappa}
\k_{\b} = \frac{\G(\frac{\b}{2}-a)\G(\frac{\b}{2}+a)\G(\frac{\b}{2}-b)\G(\frac{\b}{2}+b)}{2\pi^2\G(\b-1)\G(\b)},
\qquad
\m(w,\wb) = \Big|\frac{w-\wb}{w \wb} \Big|^{d-2} \frac{((1-w)(1-\wb))^{a+b}}{(w\wb)^2},
\ee
and the OPE data itself is the sum of $t$- and $u$-channel contributions (as in eq.~(\ref{eq:p0})):
\be
\label{eq:p52}
c(J,\D)=c^t(J,\D) + (-1)^J c^u(J,\D).
\ee 
The $u$-channel contribution may be obtained by applying the integral (\ref{eq:p2}) to the correlator
with operators 1 and 2 swapped.

Notice that the conformal block $G_{\D+1-d,J+d-1}(w,\wb)$ appearing in the inversion formula above is not the usual block,
it has the roles of $J$ and $\D$ reversed; we may call it the ``inverted block". (This reversal is a Weyl reflection of the so$(d,2)$ Lie algebra.)  This is analogous to the substitution $P_J\mapsto Q_J$ in eq.~(\ref{eq:p0}).
One can draw a close analogy between between 4-point CFT correlators
and 4-particle amplitude scattering amplitudes, see Table~\ref{f3}.

The ``${\rm dDisc}$" is primarily defined as a expectation value of
the double-commutator $-\frac{1}{2} \langle  0 | [O_2, O_3]  [O_1, O_4]  | 0 \rangle$, divided by the normalization factor
in eq.~(\ref{stripped correlator}).  It can be computed as a double discontinuity, or difference between three analytic continuations,
around the point $\zb=1$:
\bea
\label{dDisc}
 {\rm dDisc}\ [\mm{G}(\r,\bar \r)] &\equiv& \cos(\pi(a{+}b))\mm{G}(\r,\bar \r)
\nonumber\\ &&-\tfrac12e^{i\pi(a+b)} \mm{G}(\r, \bar{\r}^{-1}{-}i0)
%\\\nonumber &&
-\tfrac12e^{-i\pi(a+b)} \mm{G}(\r, \bar{\r}^{-1}{+}i0)
\eea
where we assume $0<\r,\bar{\r}<1$.  This represents a discontinuity since $\rzb$ and $\rzb^{-1}$
map onto the same cross-ratio $\zb$, 
see eq.~(\ref{rho}).

Physically, the dDisc is interpreted as an absorptive part because it represents one minus the survival probability of a certain state.
In particular it is positive-definite by unitarity, 
see section 2.2 of \cite{Caron-Huot:2017vep}.
In holographic theories, the double discontinuity effectively puts bulk propagators on-shell (as seen in specific tree and one-loop examples, see ref.~\cite{Alday:2017vkk}),
furthering the analogy with ${\rm Disc}\mathcal{M}$ and the Cutkowski rules.
The idea that it is sometimes easier to
approximate the dDisc than the correlator itself, as reviewed in introduction, motivates us
to try and reconstruct the correlator itself from this data.

%As it appears that the crude 
%\i The double discontinuity is positive $dDisc[\mm{G}(w,\wb)]\geq 0$.
%\i For large $N$ theories the double discontinuity kills the double trace contributions in the cross channel at subleading order in $1/N$.
%\i Bulk locality:  For large $N$ theories the inversion formula gives an upper bound on the strength of bulk higher derivative interactions.

\begin{table}[h!]
  \begin{center}
    \label{tab:table1}
    \begin{tabular}{l|c|c|c} % <-- Alignments: 1st column left, 2nd middle and 3rd right, with vertical lines in between
      $\mm{M}(s,t)$& $C_J(\cos \theta)$ & $Q_J(\cos \eta)$&$\mm{M}^t(s,t)= \int_{t_0}^\infty \frac{dt'}{t'-t}{\rm Disc}[\mm{M}(s,t)]$\\
         \hline
      $\mm{G}(z,\zb)$ & $F_{J,\D}(z,\zb)$&$G_{\D+1-d,J+d-1}(z,\zb)$&$\mm{G}^t(z,\zb)= \int_0^1 dw d\wb K(z,\zb,w,\wb){\rm dDisc}[\mm{G}(w,\wb)] $\\
    \end{tabular}
  \end{center}
    \caption{Analogous quantities between the 4-particle scattering amplitude (top row) and the CFT 4-point correlator (bottom row). The right most column shows the dispersion relation. \label{f3}}
\end{table}

\subsubsection{CFT dispersion relation from Lorentzian inversion formula}

Given the formula which extracts OPE data from the absorptive part (dDisc) in eq.~(\ref{eq:p2}), it is only natural
to insert it back into the OPE to obtain a dispersion relation for the correlator itself.
This is the procedure which led in subsection~\ref{ssec:amp1} to a dispersion relation for scattering amplitudes. 
We thus plug eqs.~(\ref{eq:p52}) and (\ref{eq:p2}) inside eq.~(\ref{eq:p1}):
\be
\mm{G}(z, \zb) = \mm{G}^t(z, \zb)+\mm{G}^u(z, \zb) + \mbox{(non-norm.)}
\ee 
where
\bea
\mm{G}^t(z, \zb) &=& 
\sum_{J=0}^{\infty}\int \frac{d \D}{2\pi i}  F_{J,\D}(z,\zb )   \frac{\k_{J+\D}}{4} %\ \ \ \ \ \ \ \ \ \ \ \ \ \ \ \ \ \ \ \ \ \ \ \ \ \ \ \ \ \ \ \ \ \  
\nn &&\times
\int_0^1 dw d\wb\ \m(w,\wb)\ G_{\D+1-d,J+d-1}(w,\wb)\  {\rm dDisc}[\mm{G}(w,\wb)] 
\eea 
and similarly for $\mm{G}^u(z, \zb)$.
Exchanging the order of integrals and sum then gives a dispersion relation in the form quoted in eq.~(\ref{eq:main1z}),
that is:
\be
\label{eq:poff00} \mm{G}^t(z, \zb) =  \int_0^1 dw d\wb K(z, \zb, w, \wb) {\rm dDisc}[\mm{G} (w, \wb)] ,
\ee
where the kernel is now given explicitly as:
\be
\label{eq:poff}
\boxed{
K(z, \zb, w , \wb) =  \frac{\m(w, \wb)}{8 \pi i} \sum_{J=0}^{\infty}\int_{\frac{d}{2}-i\infty}^{\frac{d}{2}+i\infty} d\D  \  \k_{J+\D}
F_{J,\D}(z,\zb )  G_{\D+1-d,J+d-1}(w,\wb). }
\ee 
This is a key formula, and the main goal of this paper will be to evaluate this kernel $K(z, \zb, w , \wb)$ explicitly.\footnote{
The kernel $K$ reported in introduction equals $K$ here upon symmetrization in $w\leftrightarrow \wb$.}
The integrand consists of the Euclidean harmonic function $F$, times the inverted block $G$ and times $\k_{\D+J}$ (the latter turns out to be crucial).

%%%%%%%%%%%%%%%%%%%%%%%%%%%%%%%%%%%%%%%%%%%%%%%%%%%%%%%%%%%%%%%

\section{Computing the CFT dispersion relation kernel}
\label{sec:disp}

In this section we analytically perform the sum-integral (\ref{eq:poff}), thus obtaining the kernel of the dispersion relation.
A few observations will simplify this endeavour:
\begin{itemize}
\item
We expect the kernel $K$ to be independent of space-time dimension, because eq.~(\ref{eq:poff00}) is a mathematical identity which should hold for any two-variable function $\mm{G}(z, \zb)$ satisfying certain analyticity properties (that are dimension-independent).
Indeed this is what happened in eq.~(\ref{eq:dvd14}) for the amplitude dispersion relation.
We will thus now set $d=2$, where the blocks are simpler, and verify in subsection~\ref{sec:4d} that the same result is obtained in $d=4$.
\item In a generic CFT, the integral (\ref{eq:p2}) only converges to the OPE data for large enough spin. Even for a unitary theory,
it may fail for $J=0$ and/or $J=1$.  It is unclear how to improve the Lorentzian inversion formula to reach these.
Our strategy will be to first glibly ignore this issue and assume convergence.
After the kernel is obtained, in the next section (see \ref{sec:subtleties}) we will extend its validity by means of a subtraction.
\item We will first perform the sum assuming identical external operator dimensions; this will require rather nontrivial identities.
We will then realize that the agreement between the $d=2$ and $d=4$ sums amount to interesting differential equations,
which will largely explain the form of the result and help attack the general case.
\end{itemize}

\subsection{Performing the $\D$ integration in $d=2$}

\def\Jsum{\ \widetilde{\sum_{J}}}

\begin{figure}[!h]
	\centering
	\includegraphics[width=160mm]{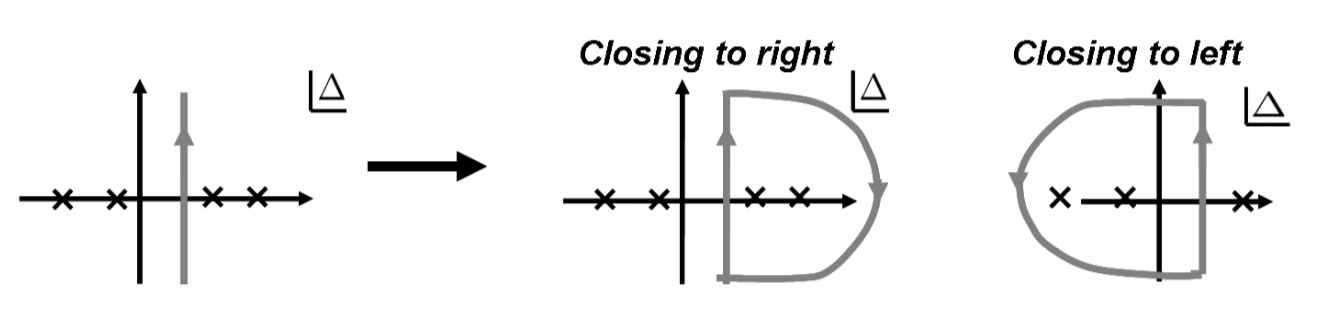}
	\caption{Left: The original integration contour of the principle series representation. One can close the contour either to the left or to the right, depending on the behaviour of the integrand at $|\D| \to \infty$. The integrand has poles on the real $\D$ axis. \label{f2}}
\end{figure}

Our first step to compute (\ref{eq:poff}) is to perform the $\D$ integral.
The idea, as shown in Fig.~\ref{f2}, is to close
the contour and use the residue theorem to get a sum over the residues of the poles.
We will need the explicit form of conformal blocks in $d=2$:
\be\begin{aligned}
G_{J,\D}(z, \zb) &= \frac{k_{\D-J}(z)k_{\D+J}(\zb)+ k_{\D+J}(z)k_{\D-J}(\zb) }{1+\d_{J,0}},
\\ k_\b(z) &\equiv  z^{\frac{\b}{2}}\  _2 F_1 (\tfrac{\b}{2}+a,\tfrac{\b}{2}+b, \b, z).
\end{aligned}\ee
Plugging into eq.~(\ref{eq:poff}), this yields two terms for the block $G(w,\wb)$, and four terms for $F(z,\zb)$ the average of
block and shadow.  We can use the $w{\leftrightarrow}\wb$ symmetry of the correlator to remove one of the former, and shadow symmetry of the other factors to neglect the shadow symmetrization, reducing the number of terms to 2:\footnote{For conciseness we define the $J$ sum with a tilde as $\widetilde{\sum}_{J} \ A_J(z, \zb) \equiv \sum^\infty_{J=0} \frac{A_J(z,\zb)}{1+\d_{J,0}}$.  %+A_J(\zb,z)
}
\be
\label{eq:p4}
K(z, \zb, w , \wb)  = \frac{\m(w, \wb)}{4 \pi i } \Jsum  \int_{1-i\infty}^{1+i\infty} d\D \ \k_{J+\D} k_{\D-J}(z)k_{\D+J}(\zb) k_{J-\D+2}(w)k_{\D+J}(\wb)  + (z{\leftrightarrow}\zb).
\ee

From now on until subsection \ref{sec:unequal} we consider the case of equal external scaling dimensions: $a=b=0$. We close the integration contour in the $\D$ plane with a semi-circle at $|\D| \to \infty$, fig.~\ref{f2}. The integrand of eq.~(\ref{eq:p4}) has the following asymptotic behaviour as $|\D| \to \infty $:
\be
\label{eq:dvd7}
\k_{J+\D} k_{\D-J}(z)k_{\D+J}(\zb) k_{J-\D+2}(w)k_{\D+J}(\wb) \to
\frac{16}{\pi} \frac{\rw ^{1+\frac{J}{2}} \rwb^{\frac{J}{2}}\rz^{-\frac{J}{2}}\rzb^{\frac{J}{2}}(\rz \rzb\rw^{-1}\rwb)^{\frac{\D}{2}}}{\sqrt{ (1-\rw^2)(1-\rwb^2)(1-\rz^2)(1-\rzb^2)}}.
\ee
%where $X=\left(\frac{(1-w)(1-\wb)}{(1-z)(1-\zb)}\right)^{(a+b)/2}$.
From this we see that when the cross-ratios are such that $\rz \rzb\rw^{-1}\rwb =1$, the $\D$ integral is divergent and the kernel will have a contact term proportional to a delta function. We compute this contact term in the next subsection.
Otherwise, the magnitude of $\rz \rzb\rw^{-1}\rwb$ determines whether we close the $\Delta$-contour to the left or to the right.
Thus we expect our kernel to contain a step function as well, ie. both ``bulk term'' and ``contact terms'' as in eq.~(\ref{KBC}), which we reproduce for convenience:
\be \label{KBC_text}
K(z, \zb,w,\wb) =K_B \theta(\rz\rzb\rwb-\rw) + K_C \frac{d\rw}{dw}\delta( \rw - \rz\rzb\rwb) + K_{\emptyset} \theta(\rw-\rz\rzb\rwb).
\ee 
The notation $K_{\emptyset}$ anticipates that the third term vanishes.

\subsubsection{The Contact term $K_C$}

Performing the $\D$ integral using the asymptotics in eq.~(\ref{eq:dvd7}) gives a delta function:
\be
\label{eq:dvd10}
\frac{1}{4\pi i}\int_{1-i \infty}^{1+i \infty}d \D  (\rw^{-1}\rz \rzb\rwb)^{\frac{\D}{2}} =\d (\rw^{-1} \rz \rzb\rwb -1).
\ee
The $J$ sum from eqs.~(\ref{eq:p4}) and (\ref{eq:dvd7}) is then simply a geometric sum:
\be
\label{eq:dvd9}
\Jsum  \rw ^{1+\frac{J}{2}} \rwb^{\frac{J}{2}}\rz ^{-\frac{J}{2}}\rzb ^{\frac{J}{2}} 
+ (z{\leftrightarrow}\zb) =
\rz\rzb\rwb \frac{1-\rz\rzb\rwb^2}{(1-\rz\rwb)(1-\rzb\rwb)}
\ee
where we have used the constraint from the $\delta$-function to eliminate $\rw$ from the result.
Combining eqs.~(\ref{eq:p4})-(\ref{eq:dvd9}) gives the result for the contact term of the kernel:
\be
 K\supset \frac{16}{\pi} \frac{1}{(w\wb)^2} \frac{\rw^2 \d (\rw- \rz \rzb\rwb)}{\sqrt{ (1-\rw^2)(1-\rwb^2)(1-\rz^2)(1-\rzb^2)}}
 \frac{1-\rz\rzb\rwb^2}{(1-\rz\rwb)(1-\rzb\rwb)}.
\ee
Dividing by the $\delta$-function and Jacobian $\frac{d\rw}{dw}$ included in eq.~(\ref{KBC_text}), this gives
the formula recorded in the introduction, namely:
\be
K_C(z,\zb,\wb) = \frac{4}{\pi} \frac{1}{\wb^2} \left(\frac{1-\rz^2\rzb^2\rwb^2}{(1-\rz^2)(1-\rzb^2)(1-\rwb^2)}\right)^{1/2}
\frac{1-\rz\rzb\rwb^2}{(1-\rz\rwb)(1-\rzb\rwb)}. \label{KC text}
\ee
The notation choice (\ref{KBC_text}), with the Jacobian factored out, allows to directly integrate out $w$, leaving a single integral
over $\wb$:
\be
\label{eq:dvd156}
\mm{G}^t(z,\zb)\Big|_{\rm contact} =  \int_0^1 d  \wb \ K_C ( z, \zb, \wb)\  {\rm dDisc}[G(w,\wb)]    \Big|_{\rw = \rz \rzb \rwb}\ .
\ee 

\subsubsection{The Bulk term $K_B$}

We now move on to compute the kernel when $\rz \rzb\rw^{-1}\rwb \neq 1$.
From eq.~(\ref{eq:dvd7}) we see that when the cross-ratios are in the regime $\rz \rzb\rw^{-1}\rwb >1$, we can close the contour to the left ({\it ie.} ${\rm Re}(\D) <1$), and the contribution from the arc at infinity will give zero. Likewise, when $ \rz \rzb\rw^{-1}\rwb <1$  we close the contour to the right  ({\it ie.} ${\rm Re}(\D) >1$) in order to drop the contribution from the arc at infinity, fig.~\ref{f2}. 

Now we use the residue theorem to compute the $\D$ integral as a sum over residues of all the poles of the integrand of eq.~(\ref{eq:p4}).
Each one of the four hypergeometric functions $k$'s has a tower of poles, and also $\k_\b$ has a tower of poles.
Performing the residue analysis, we find a few remarkable cancelations which significantly simplify the analysis. The first major simplification is that the poles of the conformal blocks always cancel in pairs after summing over $J$, and thus they give a zero contribution.  This is the same mechanism as underlies the cancellation of spurious poles in the
harmonic decomposition (\ref{eq:p1}), see \cite{Caron-Huot:2017vep,Simmons-Duffin:2017nub,Gromov:2018hut}.
Furthermore, $\k_{\D+J}$ does not have any poles on the right (see eq.~(\ref{kappa}) with $a=b=0$),
thus all the poles cancel. The kernel is identically zero in this region!
\be
K(z, \zb, w , \wb)  = 0 \qquad \text{for} \qquad \rw > \rz \rzb \rwb.
\ee
In other words the kernel is proportional to a unit step function $K(z, \zb, w , \wb)\propto \theta ( \rz \rzb\rw^{-1}\rwb-1)$.
This was expected physically, since the Lorentzian inversion formula is known to commute with the lightcone expansion:
the step function ensures that the $z\to 0$ limit of the correlator is determined by the $w\to 0$ limit of the dDisc.
 
In the kinematics in which we close to the left, the kernel is non-zero. Again the spurious poles of the conformal blocks cancel out,
but now there is a tower of double poles coming from $\k_{\D+J}$. These can be exhibited from the definition:
\be
\label{eq:dvd10a}
\k_\b \equiv \frac{1}{2\pi^2} \frac{\G^4(\frac{\b}{2})}{\G(\b)\G(\b-1)} = 
-\frac{\cot^2(\frac{\pi \beta}{2})}{\pi^2}  \frac{1}{\k_{2-\b}} .
\ee
Since $\beta=\Delta+J$, we can label the poles by a positive integer $m$:
\be
\D_{\rm pole}= -J-2m, \ \  \text{with} \ \ \  m=0,1,2, \ldots \infty.
\ee
%The coefficients of the double poles are:
%\bea
%-\frac{4}{\pi^4} \frac{1}{\k_{2m+2}} \times k_{-2J-2m}(z) k_{-2m} (\zb)  k_{2J+2m+2} (w)  k_{-2m} (\wb)     
%\eea 
Thus from eqs.~(\ref{eq:p4}), (\ref{eq:dvd10a}), and the residue theorem, we have:
\be
\label{eq:dvd11}
K_B=\frac{2}{\pi^2} \frac{1}{w^2 \wb^2}  \Jsum   \sum_{m=0}^{\infty} \frac{d}{dm'} \bigg( \frac{k_{-2J-2m'}(z) k_{-2m'} (\zb) 
 k_{2J+2m'+2} (w)  k_{-2m'} (\wb)  }{2\pi^2\k_{2m'+2}}   \bigg)  \Bigg|_{m'=m}  + (z{\leftrightarrow} \zb),
\ee
where we took the derivative with respect to $m'$ (as required by the residue theorem for the case of double poles), and then plugged the integer value $m$.
Now we notice that $J$ appears in only two hypergeometric functions; in fact the $J$-sum is telescopic and can be computed exactly,
see eq.~(\ref{eq:dvd12a}).  Performing the $J$-sum first we thus obtain
\be\label{eq:pol7}
K_B = 
\frac{4}{\pi^2}\frac{1}{w^2\wb^2}
\mm{D}_2 \sum_{m=0}^{\infty} \frac{d}{dm'} \bigg( \frac{\G^2(2m'+1)}{2\G^4(m'+1)}  k_{-2m'}(z) k_{-2m'}(\zb) k_{-2m'}(\wb) k_{2m'+2}(w)   \bigg) \Bigg|_{m'=m}     \,,
\ee
where $\mm{D}_2$ is a first-order differential operator acting on $z,\zb$ and $w$ and defined in eq.~(\ref{eq:dvd12}).
To summarize, the dispersion kernel $K$ defined by eq.~(\ref{eq:poff00}) is now written explicitly
in the form (\ref{KBC_text}) with the contact term (\ref{KC text}) plus the bulk part (\ref{eq:pol7}), the latter still to be simplified.
It remains to perform the sum $m$ over the tower of poles.
This sum seems formidable: the summand is a derivative $\frac{d}{dm'}$ of a product of 4 hypergeometric functions. Amazingly, it can be performed exactly!

\subsection{Main result from Legendre PPPQ sum}

We will now perform the sum in eq.~(\ref{eq:pol7}), namely:
\be
 S\equiv \sum_{m=0}^{\infty} \frac{d}{dm'} \bigg( \frac{\G^2(2m'+1)}{2\G^4(m'+1)}  k_{-2m'}(z) k_{-2m'}(\zb) k_{-2m'}(\wb) k_{2m'+2}(w)   \bigg) \Bigg|_{m'=m}. \label{eq:pol77}
\ee
% this can be viewed as d/dm PPPQ
To get some intuition, we first notice that, near $\wb\to 1$, each term has at most a logarithmic singularity.
This is because $k_{-2m'}(\wb)$ is polynomial for integer $m'$; a singularity can only appear when
the $\frac{d}{dm'}$ derivative acts on $k_{-2m'}(\wb)$,
 \be
 \label{eq:yu1}
\frac{d k_{-2m'}(\wb) }{dm'}\Big|_{m'=m}
= \frac{\G^2(m+1)}{\G(2m+1)} P_m(\hat{\wb}) \times \tfrac12\log(1-\wb)+ \text{(non-singular)}
\ee
where for conciseness in this section we use a hat notation in which $\hat{w}\equiv \frac{2}{w} -1$.
Let us first focus on the coefficient of the $\log$ term, that is the discontinuity around $\wb=1$.

We also notice that plugging $m'=m=\text{integer}$, the hypergeometric functions reduce to Legendre functions:
\bea
k_{-2m}( z) =  \frac{\G^2(m+1)}{\Gamma(2m+1)}  P_{m} (\hat{z}), \qquad k_{\b}( z) =2 \frac{\G(\b)}{\G^2(\frac{\b}{2})}  Q_{\frac{\b}{2}-1} (\hat{z}) .
\eea 
where $P_m(\hat{z})$ and $Q_m(\hat{z})$ are Legendre polynomials and Legendre functions of the second kind, respectively\footnote{
We pick the branch of $Q_m(x)$ which is analytic at large $m$, where it decays like $x^{-m-1}$;
we warn the reader that this is not the default branch picked by {\it e.g.} Mathematica.}.
Thus the log part of the sum becomes:
\be
\label{eq:jdpwm}
S\Big|_{\frac12\log(1-\wb)}= \sum_{m=0}^{\infty}   (2m+1) P_{m}(\hat{z})P_{m}(\hat{\zb}) P_{m}(\hat{\wb}) Q_{m}(\hat{w}).
\ee
So we must now compute this $(2m+1)P_mP_mP_mQ_m$ sum.
Luckily, the coefficient $ (2m+1)$ in the sum is the canonical coefficient which often appears with Legendre functions!
Encouragingly, we further notice that a similar sum
appeared for the scattering amplitude dispersion relation in eq.~(\ref{eq:dvd14}),
involving $(2m+1)P_mQ_m$---which can be realized in the limit $\zb,\wb\to 1$ of the current one.

It turns out that such sums (with precisely the coefficient $(2m+1)$) have been evaluated in the mathematics literature,
dating back to Watson who computed a PPP sum \cite{Watson}.
Specifically, we use the result in eqs.~(3.8)-(3.10) of \cite{baranov1},
who computed the $PPPP$ sum.
To uplift his result to our $PPPQ$ sum in eq.~(\ref{eq:jdpwm}), we need simply replace the $P_J(\hat{w})$ with a $Q_J(\hat{w})$, which can be done using the single-variable dispersion relation in the footnote below eq.~(\ref{eq:p000}).
In fact this step is completely trivial: the PPPP sum given in  \cite{baranov1} is defined in the interval $\hat{\wb}\in[-1,1]$, and has square-root branch points at the boundary.  The function whose discontinuity is this, has exactly the same functional form, but now viewed as a function of the complex plane minus the interval.
The result of the sum is thus (see also \cite{Hearingen,Rahman}):
\be \label{S log term}
 S\Big|_{\frac12\log(1-\wb)}=  \frac{4\rw \rwb \rz \rzb \textbf{K} (x) }{\pi \sqrt{(\rwb\rz- \rw \rzb)(\rwb\rzb- \rw \rz)(\rz\rzb- \rw \rwb)(1-\rw\rwb\rz\rzb)}}  \ ,
\ee
where $\textbf{K} (x) $ is the elliptic integral of the first kind:
\be
\textbf{K} (x)  \equiv \int_0^{\frac{\pi}{2}} \frac{d\theta}{\sqrt{1-x \sin^2 \theta}} = \tfrac{\pi}{2} \ {}_2F_1(\tfrac{1}{2},\tfrac{1}{2},1,x),
\ee
and $x$ is the following combination of $\r$'s, recorded previously in eq.~(\ref{x}):
\be
x \equiv  \frac{\rz \rzb \rw \rwb (1- \rz^2) (1- \rzb^2) (1- \rw^2) (1- \rwb^2)}{
(\rwb\rz- \rw \rzb)(\rwb\rzb- \rw \rz)(\rz\rzb- \rw \rwb)(1-\rw\rwb\rz\rzb)}.
\ee
We are not quite done yet---recalling eq.~(\ref{eq:pol77}), we need to account for the derivative $d/dm'$, or, equivalently,
we need to find the function whose $\log$ term is eq.~(\ref{S log term}).
This appears to be a difficult task, and so we try instead to make an educated guess. As boundary data, one can directly show that
eq.~(\ref{eq:pol77}) should be regular at $\rw\to \rz\rzb\r_{\bar {w}}$, corresponding to $x\to 1$.  The $\frac12\log (1-\wb)$ term
we have found corresponds to $\log x$ as $x\to 0$.
Our guess is to look for a second solution to the same hypergeometric differential equation, but satisfying these other boundary conditions.
In fact there is a unique candidate, which turns out to be also an elliptic function:
\be
{-}\pi \textbf{K}(1-x) =  \textbf{K}(x) \log(x) + \text{non-singular}.
\ee
This equation states that the coefficient of the log singular terms of an elliptic function is itself an elliptic function, with a changed argument.
Our educated guess, extending eq.~(\ref{S log term}), is thus:
\be  \label{correct guess}
S=  \frac{-4\rw \rwb \rz \rzb \textbf{K} (1-x ) }{\sqrt{(\rwb\rz- \rw \rzb)(\rwb\rzb- \rw \rz)(\rz\rzb- \rw \rwb)(1-\rw\rwb\rz\rzb)}} \ .
\ee
A numerical evaluation of eq.~(\ref{eq:pol77}), or its series expansion at small $\wb$,
both confirm that this ansatz is correct!

We are now done; the bulk term in the kernel is obtained as $K_B=\frac{4}{\pi^2}\frac{1}{w^2\wb^2} \mm{D}_2 S$, from eq.~(\ref{eq:pol7}).
Performing some simplifications, this gives us the form recorded in eq.~(\ref{eq:pol67}),
namely\footnote{This hypergeometric function is a linear combination of elliptic integrals of the first and second kind:
\be \label{eq:elliptic2}  {}_2 F_1 (\tfrac{1}{2},\tfrac{3}{2},2,1-x) = \frac{4}{\pi (1-x)}\Big( \textbf{K}(1-x)-\textbf{E}(1-x)\Big)\ee}:
\be \label{eq:bulk text}
K_B(z, \zb, w , \wb) = 
-\frac{1}{64\pi} \left(\frac{z \zb}{w \wb}\right)^{\frac{3}{2}}
\frac{(\wb-w)(\frac{1}{z}+\frac{1}{\zb}+\frac{1}{w}+\frac{1}{\wb}-2)}
{\big((1-z)(1-\zb)(1-w)(1-\wb)\big)^{\frac{3}{4}}}\ x^{\frac{3}{2}} {}_2 F_1 (\tfrac{1}{2},\tfrac{3}{2},2,1-x) . 
\ee
An equivalent expression, suitable for integrating with respect to $\r$-variables, is:
\be\label{nice rho}\begin{aligned}
K_B(z,\zb,w,\wb) dwd\wb &= -\frac{1}{\pi} \frac{d\rw d\rwb}{(\rw\rwb)^{3/2}}
\left( \frac{\rz \rzb}{(1-\rz^2)(1-\rzb^2)}\right)^{3/2}
\left(\frac{1}{z}+\frac{1}{\zb}+\frac{1}{w}+\frac{1}{\wb}-2\right)
\\ &\qquad\times \frac{(\rw-\rwb)(1-\rw\rwb)}{\sqrt{(1-\rw^2)(1-\rwb^2)}}x^{\frac{3}{2}} {}_2 F_1 (\tfrac{1}{2},\tfrac{3}{2},2,1-x) .
\end{aligned}\ee

%\zz{Write a physical argument as to why kernel should have a theta function.}

%The $u$-channel $\mm{G}^u(z,\zb)$ is obtained by the following replacements (it is easier to write the replacements with $u,v$ coordinates instead of $z, \zb$):
%\begin{align} {\rm dDisc}[\mm{G}(u,v)]  \mapsto {\rm dDisc}[\mm{G}(\tfrac{u}{v},\tfrac{1}{v})], \quad
%K(\rz, \rzb, \rw , \rwb)  \mapsto K(-\rz, -\rzb, \rw , \rwb) . \end{align}
%\zz{correct the first of these.}

\subsection{Match with $d=4$ and differential equation}
\label{sec:4d}

We will now similarly derive the dispersion relation in $d=4$ spacetime dimension
and show that it equals the one in $d=2$, due to interesting identities.
Since the steps are very similar, we omit details and emphasize the few changes.
The conformal blocks in $d=4$ are given by:
\bea
\label{eq:poff4f}
G_{J,\D}(z, \zb) = \frac{z \zb}{\zb -z} \Big(k_{\D-J-2}(z)k_{\D+J}(\zb) -  k_{\D+J}(z)k_{\D-J-2}(\zb) \Big).
\eea 
The extra prefactor, different measure, and shift in the argument of $k$ functions (to $\D-J-2$) lead to mild changes
in eq.~(\ref{eq:p4}):
\begin{align}
\begin{aligned} \label{d=4 sum}
&K^{(d=4)} = \frac{1}{4\pi i} \frac{z \zb}{z-\zb}  \frac{\wb-w}{w^3\wb^3}
\\
&\times  \sum_{J=0}^\infty  \int_{2-i\infty}^{2+i\infty} d\D\  \k_{J+\D} k_{\D-J-2}(z)k_{\D+J}(\zb)   k_{J-\D+4}(w)k_{\D+J}(\wb) \  + \ (z \leftrightarrow \zb).
\end{aligned}
\end{align}
As in the $d=2$ case, we close the contour in the $\D$ plane and pick up the residues of the poles, being careful with
the behavior at infinity which gives rise to contact terms.

\subsubsection{Contact term}

We first compare the contact terms, which originate from the large-$\Delta$ asymptotics
given in eq.~(\ref{eq:dvd7}).
Following the steps leading to eq.~(\ref{eq:dvd9}) we find that the kernels match due to the following identity:
\bea
\rw \frac{\wb-w}{w \wb} \frac{z\zb}{z-\zb} \sum^\infty_{J=0} \bigg( \rz^{-1} (\rw \rwb\rz^{-1}\rzb)^{\frac{J}{2}} - \rzb^{-1} (\rw \rwb\rzb^{-1}\r_{ z})^{\frac{J}{2}} \bigg) \Big|_{\rw =\rwb\rz\rzb}
\nn
=  \widetilde\sum^\infty_{J=0} \bigg(   (\rw \rwb\rz^{-1}\rzb)^{\frac{J}{2}} + (\rw \rwb\rzb^{-1}\r_{ z})^{\frac{J}{2}} \bigg) \Big|_{\rw =\rwb\rz\rzb}  
\eea
which is rather surprising but can be verified by explicit computation on both sides.
We thus find that the contact term in $d=4$ matches that of $d=2$:
\be
K^{(d=2)}_C (\rwb,\rz,\rzb) =K^{(d=4)}_C (\rwb,\rz,\rzb).
\ee

\subsubsection{Bulk term}

The agreement for the bulk term will be rather more remarkable.
Again we find that spurious poles from the blocks cancel out pairwise,
so we only need to keep the poles from $\kappa$ in eq.~(\ref{d=4 sum}), which are in the left-hand $\D$-plane.
The summation over $J$ can be performed similarly to eq.~(\ref{eq:pol7}), and leads to a \emph{different} operator acting
on the \emph{same} sum $S$ defined in eq.~(\ref{eq:pol77}):
\be \label{eq:polu56}
K_B^{(d=4)} = \frac{4}{\pi^2}\frac{1}{w^2\wb^2} \mm{D}_4 S
\ee
with
\be
\mm{D}_4 \equiv \frac{z \zb}{z-\zb} \frac{w-\wb}{w\wb} \Big(   \Big(\frac{zw(1-w)}{z-w}-\frac{\zb w(1-w)}{\zb-w} \Big)\pa_w -\frac{zw(1-z)}{z-w} \pa_z +\frac{\zb w(1-\zb)}{\zb-w} \pa_{\zb}   \Big),
\ee 
instead of $\mm{D}_2$ given in eq.~(\ref{eq:dvd12}).
Remarkably, however, it is possible to verify using the explicit form of $S$ in eq.~(\ref{correct guess})
that the two kernels agree:
\be
\Big( \mm{D}_4-\mm{D}_2\Big) S= 0.
\ee
As a result, the 4d bulk kernel is equal to the 2d one!
\be
K^{(d=4)}_B(z, \zb, w , \wb) = K^{(d=2)}_B(z, \zb, w , \wb).
\ee 
In summary, we showed that the dispersion relation is the same in $d=4$ and $d=2$. 
This strengthens our intuition that the dispersion relation should not depend on the space-time dimension $d$; it would be interesting to show this in other dimensions.

The agreement between the $d=2$ and $d=4$ kernels gives us an interesting first-order
differential equation satisfied by the $PPPQ$ sum $S$. 
Turning the logic around, we can now use this differential equation
to help determine the kernel in the general case of unequal scaling dimensions.

%as we will see shortly, explains why the complicated function of four variables $K(z,\zb,w,\wb)$ factors through a single-variable $x$ in eq.~(\ref{}).

%%%%%%%%%%%%%%%%%%%%%%%%%%%%%%%%%%%%%%%%%%%%%%%%%%%%%%%%%%%%%

\subsection{Differential equation for unequal scaling dimensions}
\label{sec:unequal}

We turn to the case of a generic 4-point correlator $\langle \mm{O}_1 \ldots \mm{O}_4 \rangle$ of scalars
with unequal scaling dimension: $a= \frac{1}{2}(\D_2 -\D_1)  \neq 0$ and $b= \frac{1}{2}(\D_3 -\D_4) \neq 0$.
We will be brief and emphasize the main points.
%The final result turns out to be more complicated than the $a=b=0$ case.
%(Although certain cases, such as $a=0$, $b=\frac{1}{2}$, turn out to be simpler. We discuss these cases in the next subsection.)
There is formally no change to eq.~(\ref{eq:p4}), namely:
\be
\label{eq:uneq99}
K^{(a,b)}  = \frac{\m(w, \wb)^{(a,b)}}{4 \pi i }\  \Jsum  \int_{1-i\infty}^{1+i\infty} d\D\ 
\k_{J+\D}^{(a,b)} k_{\D-J}^{(a,b)}(z)k_{\D+J}^{(a,b)}(\zb) k_{J-\D+2}^{(a,b)}(w)k_{\D+J}^{(a,b)}(\wb) +(z{\leftrightarrow}\zb),
\ee
where we have simply made explicit the dependence on $a$ and $b$ of the various factors.
One may easily derive the contact term, by making a simple replacement in eq.~(\ref{KC text}):
\be
\label{KC ab}
K_C^{(a,b)}( z, \zb,w,\wb)  = \left(\frac{(1-w)(1-\wb)}{(1-z)(1-\zb)}\right)^{\frac{a+b}{2}} K_C^{(0,0)}( z, \zb,w,\wb) .
\ee
The bulk term comes from poles of $\k_{\D+J}$, since
spurious poles from the conformal blocks cancel in pairs just as in the $a=b=0$ case.
As opposed to that case, however, the poles of $\k_{\D+J}$ are now single poles instead of double poles.
After performing the $J$ sum using the identity in eq.~(\ref{eq:dvd17}),
we find the generalization of eq.~(\ref{eq:pol7}):
\be
\label{eq:dvd16}
K_B^{(a,b)}(z, \zb, w , \wb)  =  \m^{(a,b)}(w, \wb)\ \mm{D}_2
\Big(S^{(a,b)}_{a} +S^{(a,b)}_{-a}+S^{(b,a)}_{b}+S^{(b,a)}_{-b}\Big)
\ee
where $\mm{D}_2$, given in eq.~(\ref{eq:dvd12}), is the \emph{same} differential operator as before, and we have defined:
\be
\label{eq:polu9}
%S_{a'}^{(a,b)} \equiv \sum_{m=0}^\infty \frac{\sin(2\pi a')\G^2_{1+2a'+2m}}{2\pi m!\G_{1+2a'+m}\G_{1+a'-b+m}\G_{1+a'+b+m}}
%k_{-2m-2a'}^{(a,b)}(z) k_{-2m-2a'}^{(a,b)}(\zb) k_{-2m-2a'}^{(a,b)}(\wb)k_{2m+2+2a'}^{(a,b)}(w)
S_{a'}^{(a,b)} \equiv \sum_{m=0}^\infty \frac{\sin(2\pi a')}{2\pi\ m!}
\frac{\G^2_{1+2a'+2m}k_{-2m-2a'}^{(a,b)}(z) k_{-2m-2a'}^{(a,b)}(\zb) k_{-2m-2a'}^{(a,b)}(\wb)k_{2m+2+2a'}^{(a,b)}(w)}
{\G_{1+2a'+m}\G_{1+a'-b+m}\G_{1+a'+b+m}\sin(\pi(a-b))\sin(\pi(a+b))}
\ee
using the notation $\G_x \equiv\G(x)$. The sum in eq.~(\ref{eq:polu9}) contains products of four hypergeometric functions which cannot be reduced to Legendre functions.
Thus it may seem hopeless to try to compute it directly. However, we may say a lot about the result using differential equations.
%\zz{Note: without the $\sin$'s this is equal to the variable pppq in the mathematica file.}

A key observation is that dimension-independence still holds, that is:
\begin{align}
 (\mathcal{D}_2-\mathcal{D}_4)S^{(a,b)}_{a'}=0. \label{D2 minus D4 ab}
\end{align}
We could prove this using hypergeometric identities to rewrite the derivatives as shift on the index $m$ of the $k$ functions,
and showing that $m$ sum becomes telescopic; it may also be readily verified order by order in $w$.
Notice that both $\mathcal{D}_2$ and $\mathcal{D}_4$ are first-order differential operators (and independent of $a$ and $b$).
In fact, thanks to the manifest permutation symmetry of
 $S^{(a,b)}_{a'}$ in $(z,\zb,\wb)$, this identity and its permutations
 give \emph{two} linearly independent differential equations.
The fact that a function is annihilated by two first-order equations implies that it factors through
the two variables which represent its zero-modes, up to an overall factor:
\begin{align} \label{Sa stripped}
S_{a'}^{(a,b)} = \frac{\sqrt {z \bar {z} w\bar {w}}}{y^{1/2+a+b}} \times \tilde{S}_{a'}^{(a,b)} (x,y)
\end{align}
where $x$ is in eq.~(\ref{x}), reproduced here for convenience, and $y$ is:
\begin{align}
x &= \frac{\rz\rzb\rw\rwb(1-\rz^2)(1-\rzb^2)(1-\rw^2)(1-\rwb^2)}{(\rwb\rzb- \rw \rz)(\rwb\rz- \rw \rzb)(\rzb \rz-\rwb\rw )(1-\rw \rz \rwb\rzb)}\ ,
\\
y &= \frac{(1-\rho_{z})(1-\rho_{\zb})(1-\rho_{w})(1-\rho_{\wb})}
{(1+\rho_{z})(1+\rho_{\zb})(1+\rho_{w})(1+\rho_{\wb})} = \sqrt{(1-z)(1-\zb)(1-w)(1-\wb)}\ .
\end{align}
The sums $\tilde{S}^{(a,b)}_{a'}$ are further constrained by \emph{second-order} differential equations,
which encode that the SL(2,R) Casimir eigenvalue with respect to each of the four variable are the same
as can be seen from eq.~(\ref{eq:polu9}).
From these we find two equations on $\tilde{S}$:
\be\begin{aligned} \label{diff eqs S}
0&=\left[ x^2 (1 - x) \partial_x^2- x^2 \partial_x + y^2 \partial_y^2 + y\partial_y- 
 \tfrac12 x^2 (1 - y^2) \partial_{x}\partial_y + (\tfrac14-a^2 -b^2)\right]\tilde{S}^{(a,b)}_{a'}, \\
0&= \left[ (x^2 (1 - y)^2 + 4 x y) \partial_x\partial_y - 2y\partial_y - 4 a b\right]\tilde{S}^{(a,b)}_{a'}.
\end{aligned}\ee
These two, together with the boundary condition that
$S_{a'}^{(a,b)}(x,y)\propto x^{1/2+a'}(1+O(x))$ as $x\to 0$,
with a constant easily determined from the $m=0$ term in eq.~(\ref{eq:polu9}), completely determine the functions $S$.

Before discussing solutions, let us make an observation about the $a=b=0$ case:
the second equation can be used to fix the $y$ dependence of each term recursively in a series in $x$; when $ab=0$, it implies that the solution is independent of $y$: $\partial_y \tilde{S}^{(a,b)}_{a'}=0$. The first equation then reduces to that satisfied by the elliptic function $\sqrt{x}{\bf K}(1-x)$.
With this method it is thus straightforward to derive the result (\ref{correct guess}) which we previously only guessed.
The key is the identity in eq.~(\ref{D2 minus D4 ab}), which states that the kernels in $d=2$ and $d=4$ are the same
and which leads to eq.~(\ref{Sa stripped}).

%In the general case, with $ab\neq 0$, %the first equation can be solved explicitly when $y=1$, providing useful boundary data:
%\be  x^(1/2 + a) {}_4F_3[1/2 + a, 1/2 + a, 1 + a, 1 + a}, {1 + 2 a,    1 + a - b, 1 + a + b}, x] \ee
%the dependence on $y$ is however nontrivial and seemingly complicated.
%For example, for the first two terms we find \simon{should be some overall $x^{1/2+a'}$ ?}
%\be
%S_{a}(x,y) = 1 + x \left(\frac{(1 + 2 a)}{4} + \frac{(1 + 2 a) b}{8 (1 + a - b) y} - \frac{(1 + 2 a) b y}{8 (1 + a + b)}\right) + O(x^2),
%\ee
%and as we go deeper into the series, we find coefficients that are high-degree polynomials in $a,b$.
Instead of looking at the individual sums $\tilde{S}_{a'}^{(a,b)}$ we now focus on the specific combination in eq.~(\ref{eq:dvd16})
and the actual kernel. It is convenient to explicitly act with the differential operator $\mathcal{D}_2$ on the prefactor in
eq.~(\ref{Sa stripped}). In a convenient normalization the kernel is then
\be
 K_B^{(a,b)} =-\frac{1}{64\pi} \left(\frac{z \zb}{w \wb}\right)^{\frac{3}{2}}
\frac{(\wb-w)(\frac{1}{z}+\frac{1}{\zb}+\frac{1}{w}+\frac{1}{\wb}-2)}
{y^{3/2+a+b}}\ \tilde{K}_B^{(a,b)}(x,y) \label{KB ab}
\ee
where
\be
\tilde{K}_B^{(a,b)}(x,y)\equiv -8\pi x^2\partial_x\Big(
\tilde{S}^{(a,b)}_{a} +\tilde{S}^{(a,b)}_{-a}+\tilde{S}^{(b,a)}_{b}+\tilde{S}^{(b,a)}_{-b}\Big).
\ee
From eqs.~(\ref{diff eqs S}) we derive differential equations satisfied by $\tilde{K}$:
\be\begin{aligned} \label{diff eqs K}
 0&=\left[ x^2 (1 - x) \partial_x^2- 2x \partial_x + y^2 \partial_y^2 + y\partial_y - \tfrac12 x^2 (1 - y^2) \partial_{x}\partial_y 
  + (\tfrac94-a^2-b^2)\right]\tilde{K}_B^{(a,b)}(x,y), \\
0&= \left[ (x^2 (1 - y)^2 + 4 x y) \partial_x\partial_y - 6y\partial_y - 4 a b\right]\tilde{K}_B^{(a,b)}(x,y).
\end{aligned}\ee
These conditions ensure that the dispersion relation commutes with the $s$-channel quadratic Casimir.\footnote{
We found that the resulting dispersion in fact commutes with the quadratic Casimir in \emph{any} dimension. The dimension-dependence of the Casimir is a first-order differential operator which it might be interesting to relate to the constraint in eq.~(\ref{D2 minus D4 ab}).}
While we have not been able to solve these in closed form, we can state the following results:
\begin{itemize}
\item The kernel $\tilde{K}_B^{(a,b)}(x,y)$ is regular around $x=1$.  While this is not true for the individual sums in eq.~(\ref{eq:polu9}) (each has a logarithmic singularity), this is a special property of the combination in eq.~(\ref{eq:dvd16}).
The kernel is then the unique regular solution to (\ref{KB ab}) with the boundary condition
\be
 \lim_{x\to 1}\tilde{K}_B^{(a,b)}(x,y) = 1-4(a+b)^2 + 16ab \frac{y}{y+1}.
\ee
(We could get the constant by combining the eqs.~(\ref{diff eqs K})
into a single fourth-order one with no $y$-derivatives, and solving it along the $y=1$ line
in terms of ${}_4F_3$ hypergeometric functions.)
\item The limits as $y\to 0$ and $y\to\infty$ are regular, and equal to a simply generalization of eq.~(\ref{KC text}):\footnote{
The limits $x\to 0$ and $y\to 0$ (or $y\to\infty$) do not commute:
to compare with the $x\to 0$ limit given below eqs.~(\ref{diff eqs S}),
one needs to carefully cross the region $x\sim y\to 0$.}
\be\begin{aligned}
 \lim_{y\to 0} \tilde{K}_B^{(a,b)}(x,y) &= (1 - 4 (a + b)^2) x^{3/2 + a + b} {}_2F_1(\tfrac12 + a + b, \tfrac32 + a + b, 2, 1 - x), \\
 \lim_{y\to \infty} \tilde{K}_B^{(a,b)}(x,y) &= (1 - 4 (a - b)^2) x^{3/2 + a - b} {}_2F_1(\tfrac12 + a - b, \tfrac32 + a - b, 2, 1 - x).
\end{aligned}\ee
\item A Taylor series in $(1-x)$ can be obtained using just the second of eqs.~(\ref{diff eqs K}), together with the previous limits;
each term is polynomial in $\frac{y}{1+y}$.
\end{itemize}
In summary, for unequal scalar operators, the kernel takes the form in eq.~(\ref{eq:main1z}), with the contact term given
explicitly in eq.~(\ref{KC ab}), and bulk term in eq.~(\ref{KB ab}) implicitly described by the above.

Let us briefly comment on the special case: $a=0$ and $b=\frac{1}{2}$,
where the bulk term $K_B$ identically vanishes. (This could be seen directly from the lack of poles of $\kappa$ in eq.~(\ref{kappa}).)
This corresponds physically to a case where the double-discontinuity (\ref{dDisc}) is effectively a single discontinuity!\footnote{We thank Dalimil Mazac for this observation.}
Only the contact term (\ref{KC ab}) remains.
We observe also that it is free of square roots (when written in terms of $\r$'s).
(More generally, when $a$ is integer and $b$ is half-integer, the contact kernel $\tilde{K}_C ( z, \zb, \wb)$ does not contain square roots.) The dispersion relation then reduces to
\be  \label{degenerate disp}
\mm{G}^{t,(0,\frac12)}(z,\zb) = \frac{1}{\pi}\int_0^1 \frac{d\rwb}{\rwb^2}
\frac{1-\rz\rzb\rwb^2}{(1-\rz\rwb)(1-\rzb\rwb)} \frac{(1-\rw)(1-\rwb)}{(1-\rz)(1-\rzb)} 
{\rm dDisc}\big[\mm{G}(w,\wb)\big]_{\rw=\rz \rzb\rwb}.
\ee
Upon further inspection, this could be recognized as a single-variable dispersion relation of the form of section \ref{sec:amp2},
taken with fixed value of the ratio $\rwb/\rw$ and acting on a certain rescaling of the correlator.
This ratio more generally will play an important role in the next section.
% the variable $\eta=\sqrt{\rwb/\rw}$ (as discussed further in the next section), for a rescaled correlator. % $\mm{G}^{(a,b)}(z,\zb)(1-\rz)(1-\rzb)$.

%\textbf{what about the $d$ dependence in the $\m$ factor? no problem}

%%%%%%%%%%%%%%%%%%%%%%%%%%%%%%%%%%%%%%%%%%%%%%%%%%%%%%%%%%

\section{Direct proof of dispersion relation}
\label{sec:directproof}
Having now obtained its kernel, we will now prove directly that the dispersion integral (\ref{eq:main1z}) indeed reconstructs
correlators.
This may be viewed as a theorem in complex analysis, independent of the CFT origin of the formula.
This will show directly the validity of the formula in any dimension, and
will enable us to go beyond the situations where the Lorentzian inversion formula converges.

We begin by observing that the contact term and bulk term of the kernel (see eqs.~(\ref{KBC_text}), (\ref{KC text}) and (\ref{eq:bulk text})),
are not independent, disparate entities. Rather, they combine into the discontinuity of a single ``pre-kernel'':
\be \label{Kpre}
K_{\rm pre}(z, \zb, w , \wb) = -\frac{1}{32} \frac{\wb -w}{w \wb}  \frac{(z \zb w \wb)^{\frac{3}{2}} (\frac{1}{z}+\frac{1}{\zb}+\frac{1}{w}+\frac{1}{\wb}-2)}{((1-w)(1-\wb)(1-z)(1-\zb))^{\frac{3}{4}}} x^{\frac{3}{2}}\  _2 F_1 (\tfrac{1}{2},\tfrac{3}{2},1,x).
\ee 
Near $x=1$, the hypergeometric function above satisfies:
\be \label{pre x1}
\  _2 F_1 (\tfrac{1}{2},\tfrac{3}{2},1,x) = \frac{2}{\pi (1-x)} - \frac{1}{2\pi}\ {}_2 F_1 (\tfrac{1}{2},\tfrac{3}{2},2,1-x)  \log(1-x)
+ \text{(non-singular)}.
\ee 
Using that
\be
 1-x = \frac{(\rz- \rzb\rw\rwb)(\rzb- \rz\rw\rwb)(\rwb- \rw \rz\rzb)(\rz\rzb\rwb -\rw)}
 {(\rzb\rwb- \rw \rz)(\rz\rwb- \rw \rzb)(\rz\rzb -\rw\rwb)(1-\rw \rz \rwb\rzb)} , \label{omx}
\ee
we see that the pre-kernel has both a pole and branch cut at $\rw \to \rz \rzb\rwb$,
whose residue and discontinuity precisely match, respectively, the contact term and bulk terms:
\be
K_C = {\rm Res}_{\rw = \rz \rzb\rwb}[K_{\rm pre}], \qquad
K_B = \frac{1}{2\pi}{\rm Disc}_{\rw \to \rz \rzb\rwb} [K_{\rm pre}] \quad (\rw<\rz\rzb\rwb).
\ee
This enables to combine these terms into a single contour integral:
\be
\boxed{
  \mm{G}^t(z,\zb) = \frac{1}{2\pi i} \int_{\mathcal{C}_w} dw \int_0^1 d\wb\ K_{\rm pre}(z, \zb, w , \wb)\ {\rm dDisc}[\mm{G}(w,\wb)]}
   \label{G combined}
\ee
where $\mathcal{C}_w$ is a ``keyhole'' contour going from the origin to the origin counter-clockwise
around its maximum $w_{\rm max}$ corresponding to $\rw=\rz\rzb\rwb$ (similar to the contour $\mathcal{C}_\sigma$ in fig.~~\ref{contours}(a)).
The existence of such a pre-kernel is very suggestive of a contour deformation argument leading to the dispersion relation.

\subsection{Contour deformation trick}

\begin{figure}[!h]
	\centering
\be\begin{array}{c@{\hspace{2mm}}c}
\raisebox{6mm}{\def\svgwidth{55mm}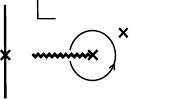}
& \hspace{10mm}\def\svgwidth{70mm}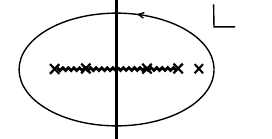
\\ (a) & (b)
\end{array}\nonumber\ee
	\caption{The integration contour used in eq.~(\ref{0 contour}) to prove the dispersion relation is a product of a keyhole and a circle.
	 (a) $\sigma_w$-plane: the keyhole $\mm{C}_\sigma$ starts and ends at $\sigma_w=0$.
	(b) $\eta_w$-plane: the integral over the circle $|\eta_w|=1$ vanishes
	and is equal to the sum of the pole and cuts in its interior.
	The pole gives minus the correlator $\mm{G}(z,\zb)$ and the cuts give the dispersion integral.
	Values of $x$ are shown in light gray above the axis.
	 \label{contours}}
\end{figure}

We will now describe a contour in two complex variables, which, fortunately for us, takes
on a simple factorized form in suitable variables.
The ``good variables'', as suggested by the degenerate case in eq.~(\ref{degenerate disp}),
are the geometric mean and ratio of $\rho$-coordinates: 
\be
 \sigma_z= \sqrt{\rz\rzb}, \quad \eta_z = \sqrt{\rz/\rzb}, \quad \sigma_w = \sqrt{\rw\rwb},\quad \eta_w = \sqrt{\rw/\rwb}.
\ee
Physically, in the Euclidean cylinder, $\sigma$ is a radial coordinate and $\eta=e^{i\theta}$ is an angular variable.
The singularities which will be relevant for our argument are shown in fig.~\ref{contours}.
The complete list of singularities of the kernel and pre-kernel come from where $x=0,1,\infty$, namely:
\be\begin{aligned}
 x&=0: & \eta_w&\in \{\pm\sigma_w,\pm\sigma_w^{-1}\}, & \sigma_w&\in \{ 0,\pm\eta_w,\pm\eta_w^{-1},\infty\}, \\
 x&=1: &\eta_w&\in \{\pm \sigma_z, \pm \sigma_z^{-1}\}, & \sigma_w&\in \{ \pm \eta_z,\pm\eta_z^{-1}\}, \\
 x&=\infty: & \eta_w&\in \{ \pm \eta_z,\pm\eta_z^{-1}\}, & \sigma_w&\in \{\pm \sigma_z, \pm \sigma_z^{-1}\}.
\end{aligned}\ee
Notice that each $\eta_w$-plane singularity is reflected four-fold: by $\eta\mapsto \eta^{-1}$, which
is parity  $w{\leftrightarrow}\wb$, and by $\eta\mapsto -\eta$, which interchanges the $t$ and $u$ channels
({\it ie.} swaps operators $1$ and $2$ in the four-point correlator).

%(The variable $\eta$ corresponds to the variable $w$ used in the contour deformation derivation of the Lorentzian inversion formula

We will now see that the dispersion relation can be derived starting from the identity:
\be
 0= \oint_{\mathcal{C}_\sigma \times \{|\eta_w|=1\}} K_{\rm pre}(z,\zb,w,\wb) G(w,\wb)
 \label{0 contour}
\ee
where the original contour, shown in fig.~\ref{contours}, is a product of a keyhole in $\sigma_w$ (similar to eq.~(\ref{G combined})),
times the unit circle $\eta_w=1$, and then deforming the contour.\footnote{
The variable $\eta_w$ was called $w$ in ref.~\cite{Caron-Huot:2017vep}
and a similar contour deformation starting from the
unit circle $|\eta_w|=1$ was used there to derive the Lorentzian inversion formula.}
This will hinge on several properties
that the pre-kernel (\ref{Kpre}) (remarkably!) combines:
\begin{enumerate}
\item It is odd under $w{\leftrightarrow}\wb$ due to the factor $(w-\wb)$.
\item The branch cut at $x=\infty$ is only logarithmic.
\item It has a simultaneous double pole when $(\sigma_w,\eta_w)=(\sigma_z,\eta_z)$, e.g. $\frac{1}{(\eta_w-\eta_z)(\sigma_w-\sigma_z)}$.
\item An analogous pole at $\eta_w=-\eta_z$ is canceled by the factor $(\frac{1}{z}+\frac{1}{\zb}+\frac{1}{w}+\frac{1}{\wb}-2)$.
\item The pre-kernel is symmetrical under $\rwb\mapsto \rwb^{-1}$ ($x\mapsto \frac{x}{x-1}$) in the region $x<1$.
This symmetry survives for the average of the two branch choices after going around $x=\infty$.
\end{enumerate}
Notice that each factor in the pre-kernel (\ref{Kpre}) has some role to play.

The vanishing of the integral along the unit circle $|\eta_w|=1$ (\ref{0 contour}) is basically due to symmetry property 1.
This is valid for generic $0<z<\zb<1$. 
Note however that property 2 is also implicitly used here, since at fixed $\sigma_w$ the unit circle contour would not be well-defined due to a branch cut
at $x=\infty$. However, thanks to property 2, the discontinuity across that cut cancels when integrated along the $\sigma_w$ keyhole.
Only the \emph{two-dimensional} contour is well-defined.

The trick now is to deform the $\eta_w$ contour inward from the unit circle.
Property 3 ensures there is a pole at $\eta_w=\eta_z$, with residue $-\mm{G}(z,\zb)$.
There is no branch cut at this point (thanks to point 2, ie. there is an expansion similar to (\ref{pre x1}) around $x\to\infty$,
and we are already taking a discontinuity in $\sigma$),
so we can keep shrinking the contour until it hits the cut at the
smaller radius $\eta_w=\sigma_z$.

In doing so, one might worry about a reflected pole at  $\eta_w=-\eta_z$, denoted by
a circle in fig.~\ref{contours}(b). Its residue would be the $u$-channel correlator $\mm{G}(z/(z-1),\zb/(\zb-1))$.
Property 4 ensures that this undesired pole does not contribute, since the numerator of the pre-kernel (\ref{Kpre})
vanishes when $(w,\wb) = (\frac{z}{z-1},\frac{\zb}{\zb-1})$:
\be
\left(\frac{1}{z}+\frac{1}{\zb}+\frac{1}{w}+\frac{1}{\wb}-2\right) = \left(-\frac{z-1}{z}-\frac{\zb-1}{\zb} +\frac{1}{w} +\frac{1}{\wb}\right).
\ee
This explains the role of this mysterious factor!

It remains to show that the cut organizes into dDisc's. We organize the cut into four segments.
On the positive axis there is the Regge region $(0,\sigma_w)$ and the Euclidean region $(\sigma_w,\sigma_z)$.
They connect at $\eta_w=\sigma_w$ or $\rwb=1$, where a lightcone is crossed ($x_{14}^2x_{23}^2=0$).
Each has a $u$-channel reflection on the negative axis.
In the Regge region we have $0<\rw<1<\rwb$ with the constraint $\rw\rwb<\rz\rzb$.
To map it to our reference region (inside the unit square) we simply need to use the symmetry under inversion $\rwb\mapsto\rwb^{-1}$ in property 5
(which interchanges $\eta_w$ and $\sigma_w$):\footnote{
It may be verified through the hypergeometric identity, valid for $x<1$:
${}_2F_1(\tfrac12,\tfrac32,1,x)=(1-x)^{3/2}{}_2F_1(\tfrac12,\tfrac32,1,\tfrac{x}{x-1})$.}
to write it to a form similar to eq.~(\ref{G combined})
\be \label{regge to Eucl}
{\rm eq.}~(\ref{0 contour})\supset
\oint_{\mm{C}_w} \int_0^1 d\wb K_{\rm pre} \times \mm{G}(\rw,\rwb^{-1}{-}i0).
\ee
where $\mm{C}_w$ is the keyhole covering $0<\rw<\rz \rzb \rwb$.
Notice that there is a Regge region above the axis, and one below the axis.
The kernel is identical in both (because its branch point at infinity is only logarithmic), so the two sides of the $t$-channel region
simply replace:
\be
\mm{G}(\rw,\rwb^{-1}{-}i0)\mapsto \mm{G}(\rw,\rwb^{-1}{-}i0)+\mm{G}(\rw,\rwb^{-1}{+}i0) \equiv
\mm{G}^{\circlearrowleft}+\mm{G}^{\circlearrowright}
\ee
where the notation emphasizes that we have gone a full circle around $\wb=1$ in the original cross-ratios.
The Euclidean region can similarly be combined, however in this case we do not need to change variables
but we need to use a symmetry of the kernel.
A subtlety is that the pre-kernel (\ref{Kpre}) (after $x$ has been around $\infty$)
has a log branch point at the boundary $x=0$ between the Regge and Euclidean regions;
however, we need the average between the two sides of the real axis ($K_{\rm pre}^{\circlearrowleft}+K_{\rm pre}^{\circlearrowright}=2K_{\rm pre}$, property 5),
leaving the desired double-discontinuity:
\be\begin{aligned}
&\oint_{\mm{C}_w} \int_0^1 d\wb \left(
 K_{\rm pre} \left(\mm{G}^{\circlearrowleft}+\mm{G}^{\circlearrowright}\right) - \big(K_{\rm pre}^{\circlearrowleft}+K_{\rm pre}^{\circlearrowright}\big) \mm{G}\right)
 \\
 &=
 \oint_{\mm{C}_w} \int_0^1 d\wb  K_{\rm pre}\times
 \left( \mm{G}^{\circlearrowleft}+\mm{G}^{\circlearrowright} -2 \mm{G}\right) \propto  \oint_{\mm{C}_w} \int_0^1 d\wb K_{\rm pre}\times {\rm dDisc}[\mm{G}]\,.
\end{aligned}\ee
The cut segments on the negative real axis similarly organize into a double discontinuity around the $u$-channel limit $\wb\to\infty$.

Let us summarize. We have proved a general result on single-valued functions of two complex variables $\mm{G}(\rz,\rzb)$.
We call a function ``single-valued'' if it satisfies the following:
\begin{itemize}
\item It is analytic in a cut plane 
$C\setminus\big\{ [1,\infty)\cup (-\infty,0]\big\}$ for each variable $\rz$ and $\rzb$
\item It is devoid of branch cuts when restricted to the Euclidean region $\rzb=(\rz)^*$
\item It satisfies $\mm{G}(\rz,\rzb) = \mm{G}(\rzb,\rz)=\mm{G}(\rz^{-1},\rzb^{-1})$ in the Euclidean region.
\end{itemize}
These properties are satisfied by any CFT correlator (as reviewed in \cite{Caron-Huot:2017vep}).
(The third condition is simply because of the way the $\rho$ variables cover the $u,v$ cross-ratios.)
The Euclidean OPE limit is $(\rz,\rzb)\to (0,0)$ and the Regge limit is $(\rz,\rzb)\to (0,\infty)$
(both of which map to $(z,\zb)\to (0,0)$ but on different sheets).
Then we showed:

\vspace{6mm}
\noindent {\bf Theorem}  Let $G(\rz,\rzb)$ be a single-valued function of two complex variables,
which vanishes sufficiently fast in the Euclidean and Regge limits.
Then the function can be recovered from its double-discontinuity
\be \label{theorem}
 \mm{G}(z,\zb) = \mm{G}^t(z,\zb) + \mm{G}^u(z,\zb), \qquad \mm{G}^t(z,\zb) \equiv
 \int_0^1 dw d\wb K(z, \zb, w , \wb) {\rm dDisc}[\mm{G}(w,\wb)],
\ee
with the kernel as quoted in introduction (eq.~(\ref{KBC})).
The necessary rate of vanishing can be estimated from convergence at $\sigma_w=0$, and along the small
arc at $\eta_w=0$ which connect the $t$- and $u$-channel Regge limits in the preceding argument.
By expanding $K_{\rm pre}$ in these limits,
we find that these arcs can be ignored provided that $\mathcal{G}(z,\zb)$ vanishes faster than $(z\zb)^{1/2}$ in both limits.
(Convergence as $\wb=1$ also naively requires a singularity no worse than $(1-\wb)^{-3/4}$, however in reality this
is naturally resolved by retaining certain arcs in the contour there, shown below, and there is no real constraint there.\footnote{
The cross-channel arcs near $\wb=1$, unlike arcs near $\wb=0$, do not invalidate the physical interpretation of the formula as building on the dDisc.})
\vspace{4mm}

Viewing eq.~(\ref{theorem}) as a result in complex analysis,
rather than a result in conformal theory, will be helpful for generalizations below.

\subsubsection{Why two variables?}

The kernel in (\ref{theorem}) is quite nontrivial,
and it is interesting to ask whether a dispersion relation with a simpler kernel than could have been possible.

It is of course possible to fix one variable, say $z$, and simply reconstruct the correlator from its discontinuity in $\zb$
 using the logic of Cauchy's theorem, as usually done for amplitudes (see section \ref{sec:amp2}).
However, such a formula will not feed on the \emph{double} discontinuity, which
has a clear physical interpretation as an absorptive part.
Rather, it would feed on the correlator in regions such as $(z,\zb)\in (0,1){\times}(1,\infty)$, whose physical interpretation remain unclear to us.
We take the viewpoint that the physical goal of a ``dispersion relation'' is to reconstruct data from some kind of ``absorptive part".
One could try to repeat the process with respect to say $\zb$ to try and get a second discontinuity,
but the basic issue which we couldn't solve is that this wouldn't avoid unphysical regions.
%({\it ie.} where $z$ is in the $t$-channel range $0<z<1$ while $\zb<0$ is in the $u$-channel).
Variables $(\rz,\rzb)$ suffer from the same issue.

One might hope to get more appealing formulas by
choosing better variables, perhaps integrating over $\eta=\sqrt{\rz/\rzb}$ in fig.~\ref{contours}b with $\sigma$ fixed.
Indeed, taking $\eta$ negative does take us to the physical $u$-channel (which is why this variable was so useful above).
The issue however is that for the integrand to organize into a dDisc, there would have to be a corresponding integral where $\eta$ is fixed
and $\sigma$ is integrated over (to provide the Euclidean correlator part of the dDisc).
The formula obtained in this paper achieves this by having a two-dimensional integral over both $\eta_w$ and $\sigma_w$,
and a nontrivial symmetry when they are exchanged (see eq.~(\ref{regge to Eucl})).

\subsection{Convergence and subtractions}

We are now positioned to overcome the limiting assumptions made in section \ref{sec:disp},
and obtain a subtracted dispersion relation that is applicable in an arbitrary unitary CFT.

\subsubsection{Subtracted dispersion relation}
\label{sec:subtleties}

Let us first see how the theorem (\ref{theorem}) clarifies the non-renormalizable terms 
in the formula (\ref{G from Gtu}) quoted in the introduction.
One such mode that is generically present is the $s$-channel identity exchange, which leads to $\lim_{z,\zb\to 0}\mm{G}(z,\zb)=1$,.
This violates the assumptions of the theorem.

A solution is simply to apply the theorem to the function $[\mm{G}-1]$, which is also single-valued, has exactly the same double-discontinuity, but vanishes faster in the Euclidean OPE limit $z,\zb\to 0$.
In general, harmonic functions (single-valued combinations of blocks and their shadows) should be subtracted for each operator of dimension less than $1$
(the same as in the harmonic analysis formula (\ref{eq:p1}) with $d=2$).
This explains the non-normalizable terms in eq.~(\ref{G from Gtu}).

There remains the question of whether the function $\big[\mm{G}(z,\zb)-\mbox{(non-norm.)}\big]$
vanishes faster than $\sqrt{z\zb}$ in the Regge limit (the limit as $z,\zb\to 0$ with $\zb$ on a second sheet),
corresponding to $\sqrt{z\zb}^{1-J}$ with exchange of a spin $J=0$ excitation \cite{Costa:2012cb}.
Unitarity implies only that the correlator stays bounded, so in general this will not be the case.
This reflects the fact that the Lorentzian inversion formula (which was the starting point of the preceding section)
may fail to converge to the OPE data for spins $J\leq 1$ \cite{Caron-Huot:2017vep}.

The theorem (\ref{theorem}) offers a simple way out: apply it to a \emph{rescaled} correlator $\mm{G} u/v=\mm{G}\tfrac{z\zb}{(1-z)(1-\zb)}$.
This is similar to the amplitude subtraction in eq.~(\ref{once subtracted 1}).
Since $\mm{G}$ is bounded in both the Euclidean and Regge limits (in any unitary CFT), this rescaled correlator
vanishes like $z\zb$ in both limits, and amply satisfies the assumptions of the theorem.
Explicitly showing the $t$- and $u$-channel contributions, this gives:
\be \label{subtracted}
\boxed{\begin{aligned}
 \frac{z\zb}{(1-z)(1-\zb)}\mm{G}(z,\zb) &= \int_0^1 dw d\wb\ K(z,\zb,w,\wb)
 \ {\rm dDisc}\left[  \frac{w\wb}{(1-w)(1-\wb)} \mm{G}(w,\wb)\right] \\
 & + \int_0^1 dw d\wb\ K(\tfrac{z}{z-1},\tfrac{\zb}{\zb-1},w,\wb)
 \ {\rm dDisc}\Big[ w\wb \mm{G}'(w,\wb)\Big],
\end{aligned}}\ee
where $\mathcal{G}'$
denotes the correlator with operators $1$ and $2$ interchanged.
The \emph{subtracted dispersion relation} (\ref{subtracted})  is a main result of this paper:
it is guaranteed to converge in any unitary CFT.
Notice that the ``non-normalizable" terms are gone: the extra power of $z\zb$ has made their subtraction unwarranted (and incorrect).

The price for better convergence at $w,\wb\to 0$ is poorer convergence near the cross-channel limit $\wb\to 1$.
This is addressed shortly; the bottom line is that $1/(1{-}\wb)$ means that the dDisc operation,
which normally suppresses double-trace operators in the $t$-channel, will leave unsuppressed the
\emph{lowest} double-twist family (ie. $t$-channel operators of twist $\Delta'-J'\approx 2\Delta_{\rm ext}$).
(Out of possible other choices, we chose $\frac{u}{v}$ so that the dDisc still suppresses higher double-twists.)

\subsubsection{Keyhole contour near cross-channel singularity}

We finally address convergence near $\wb\to 1$.  A basic fact is that the original integration contour
(see fig.~\ref{contours}) does not touch that point, so there can't be any real divergence there.
Rather, the contour encircles that point, and any apparent divergence at $\wb=1$ is an artifact of incorrectly shrinking the circle to zero size.

\def\rmax{\rho_{\rm max}}

The solution is to integrate $\rwb$ over a ``keyhole'' type contour. 
We write the result in full in the case of identical external operators.
This is best done in the following variables.  First, we parametrize the integration in terms of $t$ and $\rwb$
by setting $\rw=\rz \rzb \rwb t$, so the dispersion relation becomes:
\be \label{keyhole at 1}
 \int_0^1 dw d\wb \ K(z,\zb,w,\wb)\  {\rm dDisc}[\tilde{G}(w,\wb)]
 \mapsto \int_0^1 dt \int\limits_{\mathcal{A},\mathcal{B}_+,\mathcal{B}_-}d\rwb\ J K(z,\zb,w,\wb) F\Big|_{\rw=\rz \rzb \rwb t}
% + \int_{A,B_+,B_-} d\rwb J' K_C(z,\zb,w) {\rm dDisc}[G(w,\wb)].
\ee
where $J=\rz\rzb\rwb \frac{d w}{d \rw} \frac{d \wb}{d \rwb}$
is a Jacobian for the change of variable $(w,\wb)\mapsto (t,\rw)$, and $\tilde{G}$ stand for
either of the combinations entering in eq.~(\ref{subtracted}).
Recall that the kernel is a sum of a bulk and contact part, which are supported on $0<t<1$ and $t=1$, respectively;
they have simple expressions in $\rho$-coordinates, eq.~(\ref{nice rho}).

\begin{figure}[h]
%\centering
\raisebox{-12mm}{\def\svgwidth{55mm}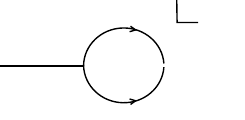}\qquad
\begin{tabular}{cccc}
region & $\rwb$ & correlator $F$ \\\hline
$\mathcal{A}$ & $\rwb\in [0,\rmax]$ & ${\rm dDisc}[\mm{G}](t,\rwb)$ \\
$\mathcal{B}_+$ & $\rwb = (\rmax)^{e^{-i\theta}}$  & $-\tfrac12 \mm{G}^\circlearrowleft(t+i0,\rwb)$ \\
$\mathcal{B}_-$ &  $\rwb = (\rmax)^{e^{i\theta}}$ & $-\tfrac12 \mm{G}^\circlearrowright(t-i0,\rwb)$
\end{tabular}
\caption{Keyhole contour in the $\rwb$ variable to avoid cross-channel singularity (at $\wb=1$). The angle runs over $\theta\in [0,\pi]$.
The radius of the circle shrinks as $\rho_{\rm max}\to 1$.}
\label{fig:keyhole}
\end{figure}

If regularization were not needed, the $\rwb$ contour would be simply the interval $[0,1]$, and the correlator
$F={\rm dDisc}[\tilde{G}]$ evaluated for $w,\wb$ given in terms of $t,\rwb$.
Keeping the full key-hole contour, it is instead the sum of a regulated interval $[0,\rmax]$ and two half-circles $[\rmax,\rmax^{-1}]$.
Explicit parametrizations and corresponding integrands are shown in figure \ref{fig:keyhole}.

The $-1/2$'s in the formula originate from the dDisc, which we recall
(see eq.~(\ref{dDisc})) for identical operators (the case considered here) is
${\rm dDisc}[\mm{G}] = \mm{G} - \tfrac12 \mm{G}^\circlearrowleft-\tfrac12\mm{G}^\circlearrowright$.
Validity of the formula require that the contour not enclose the poles at $\rwb=1/(\rz\sqrt{t})$ and $1/(\rzb\sqrt{t})$;
for real $\rz,\rzb$ this is simply achieved by requiring $\rho_{\rm max} \geq \max(\rz,\rzb)$.
In practice, in numerical examples below we chose $\rho_{\rm max}=0.9$ (adequate for $z,\zb<0.997$),
and we verified that the integral is independent of $\rho_{\rm max}$. 
The $t\pm i0$ notation indicates that the $t$-contour must avoid a branch point on the real axis (at $t=\rmax^2$).\footnote{
In practice, we parametrize $t\pm i0 \equiv \tau \pm i\epsilon \tau(1-\tau)$ where $\tau\in [0,1]$ is a real integration variable.
The offset $\epsilon$ doesn't need to be small, and we used $\epsilon=1$ in all numerical examples.}
(We note that the integrand is \emph{not} analytic at the point $\rwb=\rmax^{-1}$ where $\mathcal{B}_\pm$ meet.
At this point, the integrand matches onto the Euclidean correlator part of dDisc$[\mathcal{G}]$ at $\rwb=\rmax$.)
We find that the keyhole integral is quite practical numerically.

\section{Checks and discussion}
\label{sec:checks}

In this section we illustrate various checks and possible applications of the formula.

\subsection{Numerical check for generalized free fields}

A first sanity check is to compare both sides of the dispersion relation in the simple example of generalized free field. We consider:
\begin{align}
\label{eq:po4}
\mm{G}(w, \wb) = u^{p_1}v^{p_2} =(w \wb)^{p_1} ((1-w)(1- \wb))^{p_2}
\end{align}
Then taking the double discontinuity (for non-integer exponents),
gives for the $t$ and $u$ channel contributions respectively:
\begin{align}
&{\rm dDisc}_t [\mm{G}(u,v)] =  2 \sin^2 (\pi p_2 ) (w \wb)^{p_1} ((1-w)(1- \wb))^{p_2},
\nn
&{\rm dDisc}_t [\mm{G}(u/v, 1/v)] =  2 \sin^2 (\pi (p_1+p_2 )) (w \wb)^{p_1} ((1-w)(1- \wb))^{-p_1-p_2}.
\end{align}
Now we can plug this on the right hand side of Eq.~(\ref{theorem}).
There is a nonempty range of $p_1$ and $p_2$ for which that formula converges without subtlety, namely:
$p_1>\frac12$ and $-\frac34<p_2<\frac34-p_1$.
Computing numerically these integrals for various values of $p_1$ and $p_2$ in this range (and various values of $z,\zb$) we found
perfect match with the LHS of eq.~\ref{theorem}! 

If we relax the condition on $p_1$, we need the subtracted dispersion relation (\ref{subtracted}). 
And if $p_2$ is such that convergence is not satisfied at 1 (or $\infty$), we need to use the keyhole contour in figure~\ref{fig:keyhole}.
Again we find perfect agreement, for example when $p_1=p_2=0$, or
$p_2=2.25$ with either $p_1=1$ or $p_1=2.25$ (in the later case there is only the $t$-channel cut).
In particular, the first test confirms that the subtracted dispersion relation (\ref{subtracted}) correctly reconstructs
even the identity exchange, from the dDisc of the correlators times $u/v$.

For unequal scaling dimensions,  we did not attempt numerics because we do not have a closed form for the kernel (\ref{KB ab}).
However, we performed numerical tests in the special case $a=0$ and $b=\frac{1}{2}$ using eq.~(\ref{degenerate disp}),
and also found perfect agreement.

\subsection{Holographic correlators}
%$\mm{N}=4$ SYM correlator in holographic limit}

\begin{figure}[!h]
	\centering
	\includegraphics[width=50mm]{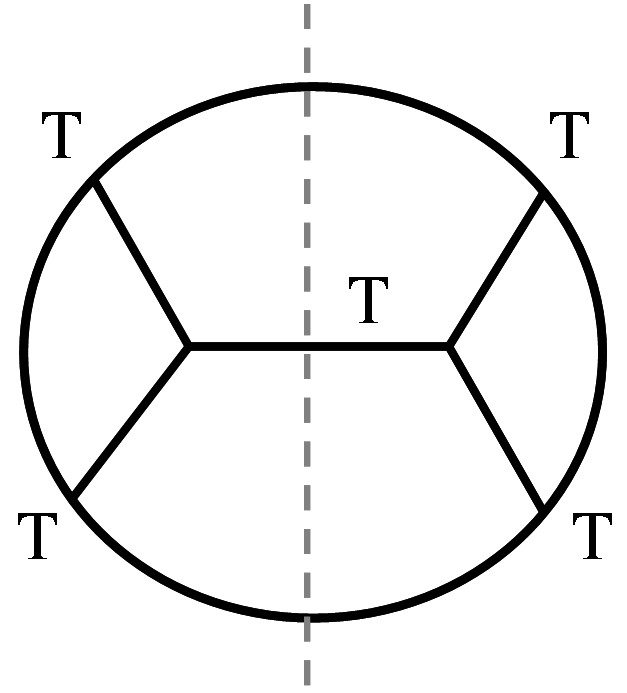}
	\caption{ The tree-level exchange diagram for the $\mm{N}=4$ stress tensor multiplet.\label{t1}}
\end{figure}

The double-discontinuity is particularly simple to compute in holographic theories, as it is saturated at tree-level by
exchange of a finite number of light single-trace operators. In some sense the dispersion relation give novel ``closed form'' expressions
for tree-level holographic correlators as an integral over conformal blocks.

For illustration, consider the correlator of stress-tensor multiplets in planar $\mathcal{N}=4$ at large 't Hooft coupling,
dual to tree-level gravity in AdS${}_5 \times S_5$. We follow the notation of \cite{Alday:2017vkk}.
(In short: because of supersymmetry, the stress tensor lies in a supermultiplet which includes scalars of dimensions
$\Delta_i=2$ in the $[0,2,0]$ representation of the SU(4)${}_R$ global symmetry.
We study the correlation functions of these scalars, projected onto the $s$-channel $[0,4,0]$ representation,
which is known to determine all other representations.  This projection has good high-energy behavior,
allowing to use the unsubtracted dispersion relation.)
The double-discontinuity in this limit is saturated by exchange of a single $t$-channel (super)conformal block
corresponding to the stress tensor multiplet, or graviton exchange in the bulk (fig.~\ref{t1});
it admits a particularly concise form \cite{Aprile:2017bgs,Alday:2017vkk}:
\be
{\rm dDisc}[\mm{G}(w,\wb)]^{(1)} = {\rm dDisc}\left[\frac{1}{1-\wb}\right]\times \left(\frac{w^2-4w^3+3w^4-2w^4\log w}{(1-w)^4} \equiv X(w)\right).
\ee
The correlator itself is the only single-value function with this double-discontinuity (and correct Regge behavior)
and is given as \cite{DHoker:1999jke,Arutyunov:2000py} (see \cite{Dolan:2000ut} for the $\bar{D}$ functions):
\be\begin{aligned}
 \mm{G}(z,\zb)^{(1)} &= \frac{u^2}{v} -u^4\bar{D}_{2422}(u,v), \label{analytic N=4}
%where $\bar{D}_{2422}$ is the special function (see \cite{Dolan:2000ut})
%\be
\\ \bar{D}_{2422} &= \pa_u \pa_v (1+u\pa_v+v\pa_v)\Phi^{(1)}(u,v),
\\ \Phi^{(1)}(u,v)&= \frac{2{\rm Li}_2(z)-2{\rm Li}_2(\zb)+ \log [\frac{1-z}{1-\zb}]\log(z\zb)}{(z-\zb)}.
\end{aligned}\ee 
We would like to see here how the dispersion relation reconstructs (\ref{analytic N=4})
starting from the elementary dDisc given above it.

First we note that the dDisc is naively zero (no branch cut), so it is really a sort of delta-function around $\wb=1$.
This can be seen explicitly from the keyhole contour in fig.~\ref{fig:keyhole}: only the semi-circles survive.
In fact it is possible to directly integrate numerically over the semi-circles and compare (successfully) with (\ref{analytic N=4}).
Let us see how the integral could be done analytically.
In fact, the two half-circles would precisely cancel each other were it not for the fact that the kernel has a log.
So ${\rm dDisc}[1/(1-\wb)]$ is effectively $-2\pi^2$ times a sort of $\delta$-function which extracts the coefficient
of $\log(1-\wb)$ in the kernel, and eq.~(\ref{keyhole at 1}) becomes
\be
 \mathcal{G}(z,\zb)^{(1)} = \int_0^{w_{\rm max}} dw
 \frac{\big[-2u^2(1-w)(u/w+1-v)\big]+\mbox{(u-channel)}}{\big(u^2 - 2 w u (1 + v) + w^2 (1 - v)^2\big)^{3/2}}\times X(w)
\label{strong 1}
\ee
where $w_{\rm max}=\frac{u}{(1+\sqrt{v})^2}$ is where $\rw=\rz\rzb$.
Adding the $u$-channel term simply cancels the $(1-v)$ term in the numerator, and doubles the remaining $u/w$ term.
A comment is in order: the integration endpoint is a branch point of the denominator, so it appears that by shrinking the circles to get a $\delta$-function
we have created a new divergence as $w=w_{\rm max}$.  One can show that the proper treatment simply amounts to integrating $w$ itself on a keyhole,
$\int_0^{w_{\rm max}}\mapsto \frac12 \int_0^0$, as in fig.~\ref{contours}a.
The integral is then unambiguous, and can also be checked numerically (giving a nontrivial confirmation of the form of the dispersion relation).

The form of the integrand of eq.~(\ref{strong 1}) makes manifest the fact that
the integral gives a combination of dilogarithms and simpler functions ---
the most complicated part can be written in the form $\int d\log(\cdots) \log(w)$,
to which standard integration algorithm can be applied, see for example \cite{Panzer:2014caa}.
(Since the square root has two branch points with respect to $w$, one has to first go to a double-cover where the square root is gone.)
It could be interesting to use this method to help understand the functions which can appear at higher loops.
%, for example using the leading discontinuity computed in \cite{Aprile:2018efk,Caron-Huot:2018kta}.

%For analytics, we first notice that the integrand is a $v$-derivative, which simplifies things somewhat:
%\be \mathcal{G}(z,\zb)^{(1)} = \int_0^{w_{\rm max}} dw\  \pa_v \frac{-u^2(u+w(1-v))}{\sqrt{u^2-2u(1+v)w+w^2(1-v)^2}} \frac{1-4w+3w^2-2w^2\log w}{w(1-w)^3}. \ee
%If we divide by $u^4$, this could further be recognized as a simple $u$ derivative, but we do not follow this route.

%%%%%%%%%%%%%%%%%%%%%%%%%%%%%%%%%%%%%%%%%%%%%%%%%%%%%%%%%%
\subsection{An integral relation between conformal blocks}
\label{sec:intrel}

The validity of the dispersion relation predicts a 
new relation between harmonic functions and the inverted block which enters the Lorentzian inversion formula
(and accessorily rederive the latter). Recall the dispersion relation:
 \begin{align} 
\label{eq:rty6}
\mm{G}(z, \zb) = \int dw d \wb \ K(w, \wb,z, \zb)\   {\rm dDisc}[ \mm{G} (w, \wb) ]
 +(t \leftrightarrow u)
 \end{align} 
The Euclidean inversion formula, in the conventions of \cite{Caron-Huot:2017vep}, is:
 \begin{align} 
\label{eq:rty7}
c(J,\D) = N(J,\D) \int_{\rm Eucl} dz d \zb \ \m (z, \zb)\ F_{J,\D}(z, \zb)\ \mm{G}(z, \zb) 
 \end{align} 
where the normalization is
$N(J,\D) \equiv \frac{\G(J+\frac{d-2}{2})\G(J+\frac{d}{2})K_{J,\D}}{2\pi \G(J+1)\G(J+d-2)K_{J,d-\D}}$. Plugging eq.~(\ref{eq:rty6}) inside eq.~(\ref{eq:rty7}) gives:
\begin{align} 
 \begin{aligned}
 \label{eq:lkj1}
&c(J,\D) =
  N(J,\D)\int dw d \wb\ \m (w, \wb)\ {\rm dDisc}[ \mm{G} (w, \wb) ] \times \\
 &   \int dz d \zb  \frac{\m (z, \zb)}{\m (w, \wb)} F_{J,\D}(z, \zb) K(w, \wb,z, \zb) +(t \leftrightarrow u)  
 \end{aligned}
 \end{align}
where we exchanged integration orders, and multiplied and divided by $\m (w, \wb)$.
This has precisely the form of the Lorentzian inversion formula (\ref{eq:p2}): Euclidean inversion plus dispersion relation gives Lorentzian inversion.
The interesting thing is that comparison reveals the following identity:
\begin{align} 
\label{eq:poly8}
\boxed{G_{\D+1-d,J+d-1}(w, \wb) = 
\frac{4 N(J,\D)}{\k_{J+\D}  } \int_{\rm Eucl} dz d \zb \frac{\m (z, \zb)}{\m (w, \wb)} F_{J,\D}(z, \zb)   K(w, \wb,z, \zb) }
 \end{align} 
This is predicted to hold for any $d$, $J$, and $\D$.
The integration
is over the complex $z$ plane, i.e $\zb =z^*$. This equation is the analog of the relation between the Legendre polynomials of the first and second kind (See table~\ref {f3}):
\be
Q_J(x)= \frac{1}{2}\int_{-1}^{1} \frac{dyP_J(y)}{y-x}\ .
\ee
While we haven't checked this relation in the general case, in appendix \ref{app:inverted} we verify it in the special case that $d=2$ and $(a,b)=(0,\tfrac12)$.  It would be interesting to understand the relationship between this identity and the light transform of
\cite{Kravchuk:2018htv}.

\subsection{3D Ising model and analytic functionals}

In the 3D Ising model, we now present numerical tests for the correlator of four $Z_2$-odd operators
($\langle\sigma\sigma\sigma\sigma\rangle$),
and discuss a possible way to reorganize the crossing equations.

A straightforward exercise (if somewhat technical) is to numerically integrate the subtracted dispersion relation in eq.~(\ref{subtracted}),
using the OPE data tabulated in \cite{Simmons-Duffin:2016wlq} to compute ${\rm dDisc}[\mm{G}]$ as a sum over
$t$-channel blocks.  It is important conceptually that the $t$-channel OPE commutes with the dispersion relation.
The basic reason is that the interior of the integration region, $w,\wb\in (0,1)$, lies within the convergence radius of the OPE.
It is also important to be careful near endpoints which lie at the boundary of convergence \cite{Rychkov:2017tpc}.
In our case these are the collinear limit $w=0$ and Regge corner $w,\wb\to 0$.
Due to absolute convergence of the (subtracted) kernel against the full correlator, and thanks to positivity of the OPE, we expect that the operations safely commute.

\begin{figure}[!h]
	\centering
\be\begin{array}{cc}
\raisebox{3mm}{\includegraphics[width=7cm]{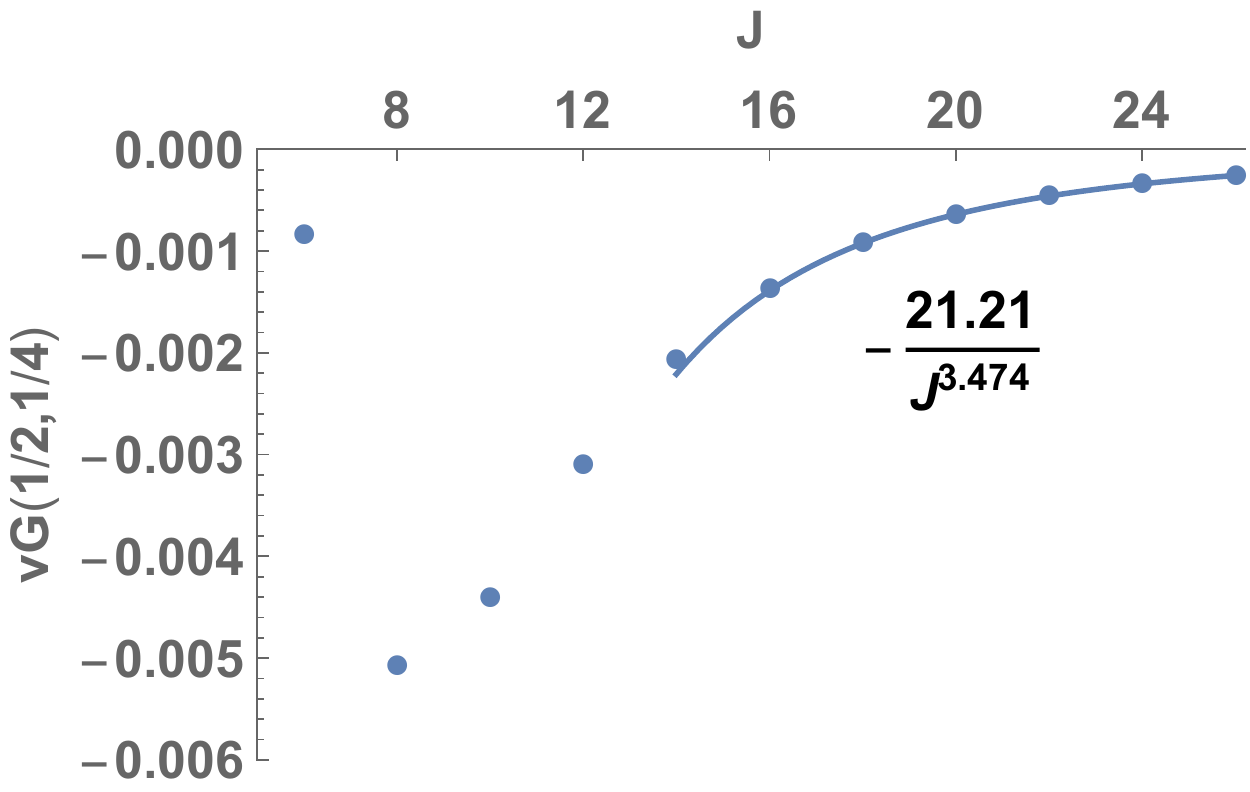}}
&
\includegraphics[width=8cm]{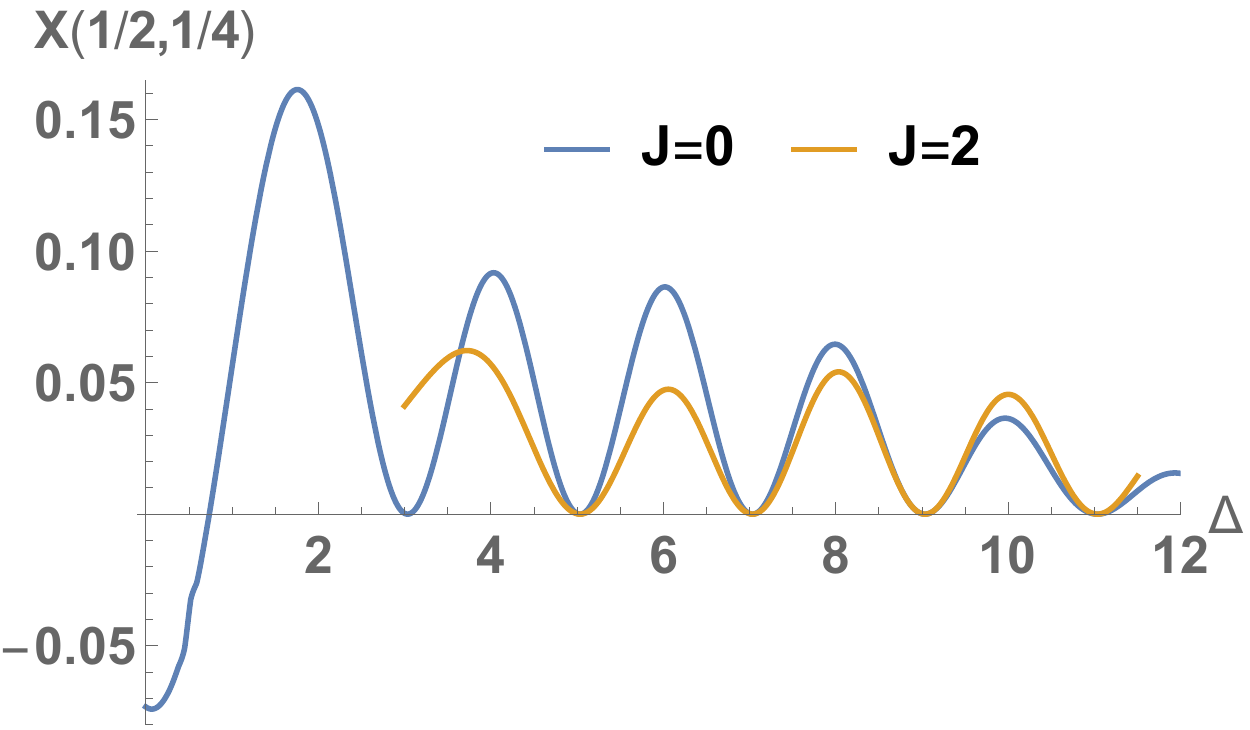} \\
\rm (a) & \rm (b)
\end{array}\nonumber\ee
\caption{(a)
High-spin tail of the contribution of the lowest-twist family $[\sigma\sigma]_{0,J}$ to the correlator in eq.~(\ref{G1}). Operators of spin $28$ and higher were resummed using the power-law fit.
(b) ``Bootstrap functional" obtained by evaluating the contribution of
spin-0 and spin-2 exchanged operators of dimension $\Delta$
to the dispersion integral $X(\tfrac12,\tfrac14)$ in eq.~(\ref{X}), with external dimensions fixed to $\Delta_\sigma$.
A rescaling envelope was applied.
Note that the curves have double zeros at \emph{non-leading} double-twist dimensions.}
 \label{functional}
\end{figure}

The most important numbers (with uncertainty in the last digit, see \cite{Simmons-Duffin:2016wlq}) are:
\be
 \Delta_\sigma = 0.518149, \qquad \Delta_\epsilon = 1.41263, \qquad f_{\sigma\sigma\epsilon} = 1.051854 \label{numbers}.
\ee
We used the 3D$\to$2D dimensional reduction formulas of ref.~\cite{Hogervorst:2016hal} to
efficiently compute the 3D conformal blocks.
Breaking the contributions into those of dominant operators and families, we find for example the correlator
at a specific point $(z,\zb)=(\tfrac12,\tfrac14)$
% {0.3687835760, 0.29684740246, 0.1649295684, 0.000300084, 0.00568071, 0.000124369, 0.000057991}
\bea\label{G1}
 v^{\Delta_\sigma} \mm{G} (\tfrac12,\tfrac14) &=&
  0.36878_{1} 
+0.29685_{\epsilon}
+ 0.16493_{T}
+0.00568_{[\sigma\sigma]_{0,\geq 4}} 
+0.00030_{\epsilon'}
+0.00018
\nonumber\\ &\approx& 0.83672
\eea
where the superscript label the cross-channel operator(s), and the last term collects all
the other operators ($[\sigma\sigma]_1$ and $[\epsilon\epsilon]_0$ families) recorded in \cite{Simmons-Duffin:2016wlq}.
Some comments are in order about the lowest twist trajectory $[\sigma\sigma]_0$.  We separated the spin-2 contribution (stress-tensor $T$) from the others of spin $J\geq 4$.  
Naively, one may have expect the whole trajectory to be suppressed by a $\sin^2(\pi(\tau/2-\Delta_\sigma))$ factor,
however, as explained below eq.~(\ref{subtracted}), the subtracted dispersion relation involves ${\rm dDisc}[\mm{G}u/v]$ which
prevents that cancelation for the lowest twist (the usual suppressions still operate for $[\sigma\sigma]_1$).
As indicated in fig.~\ref{functional}a, by evaluating the contribution of these operators up to spins $O(20)$ and fitting to a power-law,
we find that we can accurately resum the trajectory (the fit in the figure used spins 20,22,24).

Performing the similar calculation at crossing-related points $(\tfrac12,\tfrac34)$ we find:
%{0.6163518242, 0.19654854092, 0.02951021866, 0.000111355, -0.00596622, 0.0000287437, 6.31281*10^-6}
\bea
 v^{\Delta_\sigma} \mm{G} (\tfrac12,\tfrac34) &=&
  0.61635_{1} 
+0.19655_{\epsilon}
+ 0.02951_{T}
- 0.00597_{[\sigma\sigma]_{0,\geq 4}} 
+0.00011_{\epsilon'}
+0.00003
\nonumber\\ &\approx& 0.83659.
\eea
The difference of these two numbers is a crossing equation, satisfied at the $10^{-4}$ level:
\be
 X(\tfrac12,\tfrac14)= -0.00013, \qquad
 X(z,\zb) \equiv \left( v^{\Delta_\sigma} \mm{G}(u,v) - u^{\Delta_\sigma} \mm{G}(v,u) \right). \label{X}
\ee
For comparison, calculating the same correlators $vG$ at these particular points using the Euclidean OPE (and same OPE coefficients),
we find a compatible value $vG=0.83657$ (with a change of $\pm 2$ in the last digit between the two crossing-related positions).
The agreement convincingly shows that the dispersion relation indeed reconstructs the correlator.
It is presently not clear whether the (small, but significant) $10^{-4}$ error in (\ref{G1}) is due to numerical integration or truncation of the spectrum.
%The main source of error is likely numerical integration; this could probably be optimized.
%Euclidean OPE to first case: 0.6015680605867668391, 0.2147368061512493405, 0.0197943266109, 0.0000769001560025997724, 0.000396401133177672339, 9.41073338641677616*10^-6

Conceptually, one may be concerned that the sensitivity to the lowest twist trajectory means that the formula requires more than the absorptive part.  However,
in practice, the lowest-twist data is particularly well understood from the Lorentzian inversion formula.
%In practice, obtaining the OPE data at such large spins is fairly inexpensive because the lowest twist trajectory
A very crude approximation (simply feeding the identity and $\epsilon$ into the inversion formula, following \cite{Albayrak:2019gnz}),
for example reproduces the OPE coefficients of \cite{Simmons-Duffin:2016wlq} to per-mil accuracy; we used this approximation in the above, for spins 12 and higher.
Conceptually, one may view the first four contributions in eq.~(\ref{G1}) as accurately parametrized (to per-mil level) simply by three parameters: $(\Delta_\sigma,\Delta_\epsilon,f_{\sigma\sigma\epsilon})$.

It is amusing to try to constraint this crude model, for example
for a given $\Delta_\sigma$ one can find $(\Delta_\epsilon,f_{\sigma\sigma\epsilon})$
which minimizes the error in the crossing relation (and in the twist of the stress tensor).
Preliminary investigations yield a curve(s) passing through the numerical bootstrap solution (\ref{numbers}),
with values of $\Delta_\epsilon$ differing by less than $\pm 0.01$ when considering different crossing equations
(we did not observe any kink). Possibly, to close the system and also fix $\Delta_\sigma$ by such methods,
one will need to consider mixed correlators.

It is interesting that the contributions of individual blocks are very different between the Euclidean OPE
and dispersion relation: the dispersion relation is \emph{not} a term-wise rewriting of the OPE.
This becomes particularly sharp if we plot the contribution to a crossing equation, say $X(\tfrac12,\tfrac14)$,
from a given cross-channel
operator.  This gives a ``bootstrap functional", shown in fig.~\ref{functional}b, which must be orthogonal to the OPE data.
That particular functional has double zeros at all (non-leading) double-twist operators, and is mostly positive (with the exception of the identity contribution, and some lowest-twist operators at high spin, not shown).  In contrast, Euclidean functionals display no such oscillatory behavior.

One can create a few more functionals of this type.
For example, $t{\leftrightarrow}u$ crossing symmetry is not manifest because of the subtraction (\ref{subtracted}),
giving a nontrivial constraint:
\be \tilde{X}(z,\zb) \equiv  \mm{G}(z,\zb) - \mm{G}\big(\tfrac{z}{z-1},\tfrac{\zb}{\zb-1}\big) =0.\ee
A special case includes the Regge limit, for example the correlator should be real
for imaginary $\rho$'s:
\be {\rm Im}\ \mm{G}(\rz=ia, \rzb=-ib)=0, \qquad (0<a<1<b \mbox{ real}).\ee
%This constraint on OPE data is presumably hard to reach from the Euclidean OPE.
Such functionals will likely not form a complete basis (all have double zeros at the double-twists, unlike some of those in \cite{Mazac:2016qev}), but it would be interesting to compare and perhaps combine them with other functionals like those found in \cite{Paulos:2019gtx,Mazac:2019shk}.

%A key question, which we leave to future work, is whether these crossing constraints suffice to independently bootstrap the OPE data.
%As a preliminary step, we tried a simple model where we parametrize the correlation functions by the 3 parameters $\{ \Delta_\sigma,\Delta_\epsilon,f_{\sigma\sigma\epsilon}\}$.

%In terms of these 3 parameters,
%we use the Lorentzian inversion formula (exchanging $1$ and $\epsilon$) to generate the
%lowest-twist trajectory $[\sigma\sigma]_0$, and then truncate the dispersion relation to $\{1,\epsilon,[\sigma\sigma]_0\}$.
%We remove one parameter ($f_{\sigma\sigma\epsilon}$) by requiring that $[\sigma,\sigma]_{0}$ passes through
%the stress-tensor twist, $\tau_{0,2}=1$. For the reference values (\ref{}), the error $X(\tfrac12,\tfrac34)$ is about $10^{-3}$.
%We then tried to minimizing the error $X(\tfrac12,\tfrac34)$ as a function of $\Delta_\sigma$ and $\Delta_\epsilon$ --- this gives us one curve $\Delta_\epsilon^*(\Delta_\sigma)$ (where $X(\tfrac12,\tfrac34)$ vanishes).
%We then get two more crossing equations by comparing the values at corresponding
%$u$-channel points $(-1,3)$ and $(-1,\tfrac{-1}{3})$, and thus two zero-error curves.  All these curves are parallel!  No kink.

\section{Conclusion}
\label{sec:discussion}

In this work we obtained a dispersion relation for four-point correlators of conformal field theories,
reconstructing them from an ``absorptive part" (double discontinuity). 
It's kernel (given in eqs.~(\ref{KBC_text}), (\ref{KC text}) and (\ref{eq:bulk text})) was found by explicitly resuming
the Lorentzian inversion formula of \cite{Caron-Huot:2017vep,Simmons-Duffin:2017nub,Kravchuk:2018htv}.
For non-equal external operators, a differential equation was obtained, eq.~(\ref{diff eqs K}).
A subtracted dispersion relation, in eq.~(\ref{subtracted}), overcomes the limitations of the inversion formula
fully reconstructs the correlators in an arbitrary (unitary) conformal field theories. 
Various tests were performed, including in holographic theories and the 3D Ising model.

The dispersion relation holds for $d\geq 2$. For $d=1$ there is only one cross ratio $z$, and thus one could expect a simpler dispersion relation. Ref.~\cite{Mazac:2018qmi} obtained a crossing symmetric inversion formula in $d=1$. The kernel in this inversion formula is quite complicated for general scaling dimensions, precisely because it needs to give rise to a crossing symmetric correlator. Combining a known $d=1$ inversion formula with the methods that we presented in this note, one can obtain a $d=1$ dispersion relation \cite{New} containing the double-discontinuity. It is also possible \cite{New} to obtain dispersion relations for boundary/defect CFTs by starting from the Lorentzian inversion formulas of \cite{Liendo:2019jpu,Mazac:2018biw}.  Investigating the flat space limit of the dispersion relation would also be interesting,
as well as comparison with momentum space approaches (for example \cite{Gillioz:2016jnn}).

In our view, the most appealing feature is that the ``absorptive part" (or dDisc) on which the dispersion relation feeds
can often be rather accurately approximated by just the simplest exchanges, as discussed below eq.~(\ref{X}).
This strongly suggests that this is the right data around which to build a systematic expansion.
Our hope is that the dispersion relation presented here will help achieve that,
as crossing symmetry can now be directly formulated as a constraint on the dDisc.
After subtracting the simplest exchanges, the remainder of the dDisc should be a small, positive, and regular function on a square ($0{<}z,\zb{<}1$). Finding how to ``close" the equations and bootstrap this function is in our view a key next question.
%obtaining practical formulas for the mixed correlators might be important in order to close the problem.

%One can attempt to iterate the dispersion relation (plug it into itself) in order to obtain a "double dispersion relation", in terms of higher discontinuities of the correlator. Presumably such a representation would be a more crossing symmetric representation than the one in this note.  It would be interesting to take the flat space limit of that dispersion relation, in order to try to recover the Mandelstam double dispersion relation.  [SCH: won't converge...]

\acknowledgments

We thank Lorenzo Di Pietro, Shota Komatsu, Petr Kravchuk, Dalimil Mazac, Joao Penedones, Balt van Rees and David Simmons-Duffin for discussions. DC thanks McGill university and Caltech for hospitality where most of this work was done.
Part of this work was done during the Bootstrap 2019 conference at Perimeter institute and supported by the Simons foundation. Work of SCH is supported by the National Science and Engineering Council of Canada, the Canada Research Chair program,
the Fonds de Recherche du Qu\'ebec - Nature et Technologies, and the Simons Collaboration on the Nonperturbative Bootstrap.
DC is supported by the European Research Council Starting Grant under grant no.\ 758903.

\appendix

\section{Identities for spin sums}
In this appendix we collect some of mathematical results that we use in the main text.
In Eq.~\ref{eq:dvd11} we used the following $J$ sum which contains 2 hypergeometric functions, which one can easily check by series expanding in $w$:
\begin{align}\label{eq:po2}
\begin{aligned}
&\sum_{J=0}^{\infty} \frac{ k_{-2J-2m'}(z)  k_{2J+2m'+2} (w)}{1+\d_{J,0}}
\\
&= \frac{zw}{z-w} \Big( k_{-2m'}(z) k_{2m'}(w) - \frac{m'^2 k_{-2m'+2}( z) k_{2m'+2}( w) }{4(4m'^2-1)} \Big) 
- \frac{ k_{-2m'}( z) k_{2m'+2}( w) }{2}
\\
&= \frac{1}{2m'+1} \left[ \frac{zw}{z-w}\Big(  (1-w)\pa_w -(1-z)\pa_z   \Big) -\frac12\right]
k_{-2m'}(z) k_{2m'+2}(w).
\end{aligned}
\end{align}
The idea behind the first equality is that $(1/z-1/w)$ times the summand on the first line can be rewritten, using hypergeometric identities, in terms of $k$-functions with shifted arguments so as to turn the sum into a telescopic one.
The second line similarly re-interpret shifted $k$-functions in terms of derivatives on simple term.
The above identity is valid for arbitrary offset $m'$, but $a=b=0$.
For non-zero $a$ and $b$, the second line form of the above identity becomes more complicated,
but the derivative form remains unchanged:
\begin{align}\label{eq:dvd17}
\begin{aligned}
&\sum_{J=0}^{\infty} \frac{ k_{-2J-2m'}^{(a,b)}(z)  k_{2J+2m'+2}^{(a,b)} (w) }{1+\d_{J,0}}
\\
&= \frac{1}{2m'+1} \left[ \frac{zw}{z-w}\Big(  (1-w)\pa_w -(1-z)\pa_z   \Big) -\frac12\right]k_{-2m'}^{(a,b)}( z) k_{2m'+2}^{(a,b)}(w).
\end{aligned}
\end{align}
Introducing the differential operator (the subscripts stands for $d=2$)\footnote{In terms of the $\r$ coordinates:
\be
\mm{D}_2 \equiv \frac{\rz \rw \Big(  (1-\rw^2)\pa_{\rw}-(1-\r^2_z)\pa_{\rz}   \Big)}{(\rz-\rw)(1-\rw \rz)} + \frac{\r_{\zb} \rw \Big(  (1-\rw^2)\pa_{\rw}-(1-\r^2_{\zb})\pa_{\rzb}   \Big)}{(\r_{\zb}-\rw)(1-\rw \r_{\zb})} -1. \nonumber
\ee}:
\be
\label{eq:dvd12}
\mm{D}_2 \equiv \frac{zw}{z-w}\Big(  (1-w)\pa_w -(1-z)\pa_z   \Big) + \frac{\zb w}{\zb-w}\Big(  (1-w)\pa_w -(1-\zb)\pa_{\zb}   \Big) -1,
\ee
this allows us to rewrite the sum (\ref{eq:dvd11}) in a concise form:
\begin{align}
\label{eq:dvd12a}
 \sum_{J=0}^{\infty} \frac{k_{-2J-2m'-2a}(z) k_{-2m'-2a}(\zb)k_{-2m'-2a}(\wb)(k_{2J+2m'+2+2a} (w) }{1+\d_{J,0}} + (z{\leftrightarrow}\zb)
 \\ =\frac{1}{2m'+1}\mm{D}_2 \Big[ k_{-2m'}(z) k_{-2m'}(\zb) k_{-2m'}(\wb) k_{2m'+2}(w)\Big].
\end{align}

\section{Inverted block from harmonic function when $a=0$, $b=\frac{1}{2}$}
\label{app:inverted}

In order to check the integral relation in Eq.~\ref{eq:poly8}, we fix $d=2$ and $a=0$, $b=\frac{1}{2}$. In this case the dispersion relation contains only the contact term $K_C$ and not the bulk term $K_B$, as we saw in Eq.~\ref{degenerate disp}. Thus Eq.~\ref{eq:poly8} becomes\footnote{Where we used $\frac{4 N(J,\D)  }{\k_{J+\D}  } =\frac{2(1+\d_{J,0})}{\pi\kappa_{J+2-\D}}$.}:
\begin{align}
\label{eq:poly8818}
\frac{ \frac{\pi}{2} \k_{J+2-\D}}{1+\d_{J,0}}    G_{\D-1,J+1}(w, \wb) =   \int_{\rm Eucl} dz d \zb \frac{\m (z, \zb)}{\m (w, \wb)}  F_{J,\D} (z,\zb)  \d (\rw-  \rwb \rz \rzb ) K_C \frac{d\rw}{dw} 
\end{align}
For $a=0$, $b=\frac{1}{2}$ the ${}_2F_1$'s of the conformal block basically simplifies to powers of $\r$:
\begin{align}
&G_{J,\D}(z,\zb)= \frac{1}{1+\d_{J,0}}(k_{\D-J}(z)k_{\D+J}(\zb)+k_{\D+J}(z)k_{\D-J}(\zb))
\nn
&= \frac{2^{2\D}  }{1+\d_{J,0}} \frac{ (\rz)^{\frac{\D-J}{2}} (\rzb)^{\frac{\D+J}{2}} + (\rz)^{\frac{\D+J}{2}} (\rzb)^{\frac{\D-J}{2}} }{(1-\rz)(1-\rzb)}
= \frac{2^{2\D}r^\D ( e^{i \theta J} + e^{-i \theta J} )}{(1+\d_{J,0}) |1-r e^{i\theta }|^{2}} 
\end{align}
where in the last equality we changed to polar coordinates $\rz = r e^{i\theta}$, and used the fact that in Euclidean we have $\bar{\r}_z =\r^*_z$. Similarly, we can write $K_C$ and the inverted block in polar coordinates. Plugging these ingredients back in Eq.~\ref{eq:poly8818}, we can now check that Eq.~\ref{eq:poly8818} gives a correct result:
\begin{align}
&\frac{1}{ \frac{\pi}{2} \k_{J+2-\D}   G_{\D-1,J+1}(w, \wb)  } \int_{\rm Eucl} dz d \zb \frac{\m (z, \zb)}{\m (w, \wb)}  F_{J,\D} (z,\zb)  \d (\rw-  \rwb \rz \rzb ) K_C \frac{d\rw}{dw}
 \nn
& =    \frac{  \rw^{\frac{5}{2}} \rwb^{-\frac{3}{2}}   (1-\rw \rwb) (\r_w \bar{\r}_w)^\frac{-J}{2} }{  (\rw)^{\frac{2-\D}{2}} (\rwb)^{\frac{\D}{2}} + (\rw)^{\frac{\D}{2}} (\rwb)^{\frac{2-\D}{2}}  }
\int_0^{2\pi} d\theta \int_0^\infty  dr   \frac{   r^{\D} +r^{2-\D } }{4\pi r^5}   \frac{  ( e^{i \theta J} + e^{-i \theta J} ) \d (r- \sqrt{\frac{\rw}{\rwb}}  ) }{   | \sqrt{\rw \rwb} -e^{-i\theta}  |^2  }
 \nn
& =  \frac{   (1-\rw \rwb)  }{4\pi   ( \r_w \bar{\r}_w)^\frac{J}{2}}  \int_0^{2\pi}    \frac{ ( e^{i \theta J} + e^{-i \theta J} ) d\theta }{  | \sqrt{\rw \rwb} -e^{-i\theta}  |^2  } 
\nn
& =  \frac{   (1-\rw \rwb)  }{4\pi i  ( \r_w \bar{\r}_w)^\frac{J}{2}} \int_{|Z|=1}   \frac{(Z^{J-1}+ Z^{-J-1})dZ}{(Z- \sqrt{\rw \rwb}) (Z^{-1}-  \sqrt{\rw \rwb}) } 
=1
\end{align}
Where in the third line we performed the $r$ integral over the delta function, in the third line wrote the remaining $\theta$ integral as a contour integral in the complex $Z\equiv e^{i\theta}$ plane, and then used the residue theorem in the complex $Z$ plane to get 1.

\bibliographystyle{JHEP}
\bibliography{Dispersion-refs}

\end{document}